\documentclass[floatfix,twocolumn,prl,amsmath]{revtex4}
\usepackage{graphicx}
\usepackage{bm}
\usepackage{amssymb}
\usepackage{dcolumn}
\usepackage{subfigure}
\usepackage{threeparttable}
\usepackage{multirow}
\usepackage{mathrsfs}
\DeclareMathAlphabet{\mathscrbf}{OMS}{mdugm}{b}{n}
\usepackage[usenames,dvipsnames]{color}
\usepackage{mathtools}

\begin{document}
\newcommand{\vn}[1]{{\boldsymbol{#1}}}
\newcommand{\vht}[1]{{\boldsymbol{#1}}}
\newcommand{\matn}[1]{{\bf{#1}}}
\newcommand{\matnht}[1]{{\boldsymbol{#1}}}
\newcommand{\bege}{\begin{equation}}
\newcommand{\ee}{\end{equation}}
\newcommand{\bal}{\begin{aligned}}
\newcommand{\defbar}{\overline}
\newcommand{\SM}{\scriptstyle}
\newcommand{\eal}{\end{aligned}}
\newcommand{\torkance}{t}
\newcommand{\udot}{\overset{.}{u}}
\newcommand{\exponential}[1]{{\exp(#1)}}
\newcommand{\phandot}[1]{\overset{\phantom{.}}{#1}}
\newcommand{\phandag}{\phantom{\dagger}}
\newcommand{\Trace}{\text{Tr}}
\newcommand{\Bxc}{\Omega}
\newcommand{\crea}[1]{{c_{#1}^{\dagger}}}
\newcommand{\annihi}[1]{{c_{#1}^{\phantom{\dagger}}}}
\newcommand{\mubo}{\mu_{\rm B}^{\phantom{B}}}
\newcommand{\rmd}{{\rm d}}
\newcommand{\magdir}{\hat{\vn{M}}}
\newcommand{\rme}{{\rm e}}
\newcommand{\intkspa}{\int\!\!\frac{\rmd^d k}{(2\pi)^d}}
\newcommand{\intkspapri}{\int\!\!\frac{\rmd^d k'}{(2\pi)^d}}
\newcommand{\gret}{G_{\vn{k} }^{\rm R}(\mathcal{E})}
\setcounter{secnumdepth}{2}
\title{Chiral damping, chiral gyromagnetism and current-induced
torques in textured one-dimensional Rashba ferromagnets}
\author{Frank Freimuth}
\email[Corresp.~author:~]{f.freimuth@fz-juelich.de}
\author{Stefan Bl\"ugel}
\author{Yuriy Mokrousov}
\affiliation{Peter Gr\"unberg Institut and Institute for Advanced Simulation,
Forschungszentrum J\"ulich and JARA, 52425 J\"ulich, Germany}
\date{\today}
\begin{abstract}
We investigate Gilbert damping, spectroscopic gyromagnetic ratio and current-induced
torques in the one-dimensional Rashba model with an additional 
noncollinear magnetic exchange field.
We find that the Gilbert damping differs between left-handed and
right-handed N\'eel-type magnetic domain walls due to the combination of 
spatial inversion asymmetry and spin-orbit interaction (SOI), consistent
with recent experimental observations of chiral damping.
Additionally, we find that also the spectroscopic $g$ factor differs between left-handed and
right-handed N\'eel-type domain walls, 
which we call chiral
gyromagnetism. 
We also investigate the gyromagnetic ratio in the Rashba model with
collinear magnetization, where we 
find that scattering corrections to the $g$ factor vanish for zero SOI,
become important for finite spin-orbit coupling, and tend to stabilize the gyromagnetic ratio
close to its nonrelativistic value.  
\end{abstract}

\maketitle
\section{Introduction}
In magnetic bilayer systems with structural 
inversion asymmetry
the energies of left-handed
and right-handed N\'eel-type domain walls differ due to the
Dzyaloshinskii-Moriya 
interaction (DMI)~\cite{dmi_moriya,dmi_dzyalo,heide_dmi_few,heide_dmi_mnw}. 
DMI is a chiral interaction, i.e.,
it distinguishes between left-handed and
right-handed spin-spirals. 
Not only the energy is sensitive to
the chirality of spin-spirals. Recently, it has been reported that
the orbital magnetic moments differ as well between left-handed and
right-handed cycloidal spin spirals in magnetic 
bilayers~\cite{com,fabiancom}.
Moreover, the experimental observation of asymmetry in the
velocity of domain walls driven by magnetic fields suggests that also the
Gilbert damping is sensitive to 
chirality~\cite{chiral_damping_magnetic_domain_walls,phenomenology_chiral_damping}.

In this work we show that additionally the spectroscopic gyromagnetic ratio $\gamma$ is
sensitive to the chirality of spin-spirals. The spectroscopic gyromagnetic 
ratio $\gamma$
can be defined by the equation
\bege\label{eq_define_gamma}
\frac{\rmd\vn{m}}{\rmd t}=
\gamma
\vn{T}
,
\ee
where $\vn{T}$ is the torque 
that acts on the magnetic moment $\vn{m}$ and
$\rmd\vn{m}/\rmd t$ is the resulting rate of change.
$\gamma$ enters the Landau-Lifshitz-Gilbert equation (LLG):
\bege
\frac{\rmd\hat{\vn{M}}}{\rmd t}=
\gamma\hat{\vn{M}}\times\vn{H}^{\rm eff}
+\vn{\alpha}^{\rm G}
\hat{\vn{M}}\times\frac{\rmd \hat{\vn{M}}}{\rmd t},
\ee
where $\hat{\vn{M}}$ is a normalized vector that points in the
direction of the magnetization and the tensor $\vn{\alpha}^{\rm G}$
describes the Gilbert damping.
The chirality of the gyromagnetic ratio provides another
mechanism for asymmetries in domain-wall motion between
left-handed and right-handed domain walls.

Not only the damping and the gyromagnetic ratio exhibit chiral
corrections in inversion asymmetric systems but also the current-induced torques.
Among these torques that act on domain-walls are the
adiabatic and nonadiabatic spin-transfer 
torques~\cite{current_induced_magnetization_dynamics_disordered_PhysRevB.74.144405,first_principles_nonadiabatic_stt_Ni_Fe,nonadiabatic_stt_real_materials,intrinsic_spin_torque_no_soc}
and the spin-orbit 
torque~\cite{matching_dw_and_sot_efficient_PhysRevB.87.020402,cit_textured_rashba_ferromagnets,ibcsoit,invsot}.
Based on phenomenological grounds additional types of torques have
been suggested~\cite{phenomenology_sot}.
Since this large number of contributions 
are difficult to disentangle experimentally,
current-driven
domain-wall motion in inversion asymmetric systems is 
not yet fully understood. 

The two-dimensional
Rashba model with an additional exchange splitting has been
used to study spintronics effects associated with the interfaces in
magnetic bilayer 
systems~\cite{manchon_zhang_2008,rashba_review,quantum_kinetic_rashba_macdonald,sots_2d_rashba_ferromagnets_titov,microscopic_theory_sot}.
Recently, interest in the role of DMI in one-dimensional magnetic chains
has been 
triggered~\cite{dmi_chiral_magnetism_zigzag_chains,dmi_Pt_step_edges_PhysRevB.94.024403}.
For example, the magnetic moments in bi-atomic Fe chains on the 
Ir surface order in a 120$^{\circ}$ spin-spiral state due to DMI~\cite{Fe_Ir_Menzel}.  
Apart from DMI, also other chiral effects, such as chiral damping and chiral gyromagnetism,
are expected to be important in one-dimensional magnetic chains on heavy metal substrates. 
The one-dimensional Rashba 
model~\cite{spin_split_bands_onedim_chain,onedim_Si_Au} 
with an additional exchange splitting
can be used 
to simulate spin-orbit driven effects in one-dimensional magnetic wires on
substrates~\cite{monoatomic_fe_co_wires_pt_surface_step_edge,influence_substrate_mae_monoatomic_nanowires_PhysRevB.73.134428,oscillatory_mae_one_dimensional_atomic_wires_PhysRevLett.93.077203}.
While the generalized Bloch theorem~\cite{generalized_bloch_theorem} 
usually cannot be used 
to treat spin-spirals when SOI is included in the calculation,
the one-dimensional Rashba model has the advantage
that it can be solved with the help of the generalized Bloch 
theorem, or with
a gauge-field approach~\cite{gauge_fields_spintronics}, 
when the spin-spiral is 
of N\'eel-type. 
When the generalized Bloch theorem cannot be employed
one needs to resort to a 
supercell approach~\cite{dmi_copt_thiaville},
use open boundary 
conditions~\cite{SOC_torque_ballistic_domain_walls_PhysRevB.93.224415,gilbert_damping_noncollinear_ferromagnets_PhysRevLett.113.266603},
or apply 
perturbation theory~\cite{fabiancom,phase_space_berry,mothedmisot,current_induced_magnetization_dynamics_disordered_PhysRevB.74.144405,geotexmosoi,microscopic_calculation_thermally_induced_stt} 
in order to study spintronics effects
in noncollinear magnets with SOI.
In the case of the one-dimensional Rashba model 
the DMI and the exchange parameters were calculated both
directly based on a gauge-field approach and  
from perturbation theory~\cite{geotexmosoi}. 
The results from the two approaches
were found to be in perfect agreement. Thus, the one-dimensional
Rashba model provides also an excellent opportunity to verify expressions
obtained from perturbation theory by comparison to the results
from the generalized Bloch theorem or from the gauge-field approach.

In this work we study chiral gyromagnetism and chiral damping 
in the one-dimensional Rashba model with
an additional noncollinear magnetic exchange field. The one-dimensional
Rashba model is very well suited to study these SOI-driven
chiral spintronics effects, because it can be solved 
in a very transparent way without the need
for a supercell approach, open boundary conditions or perturbation
theory. 
We describe scattering effects by the Gaussian scalar disorder model. 
To investigate the role of disorder for the gyromagnetic ratio in general, we
study $\gamma$ also in the two-dimensional Rashba model with collinear
magnetization.
Additionally, we compute the current-induced torques in the one-dimensional
Rashba model.

This paper is structured as follows:
In section~\ref{sec_rashba_onedim} we introduce the
one-dimensional Rashba model.
In section~\ref{sec_gilbert_and_gyromagnetic_formalism} we 
discuss the formalism for the calculation of the Gilbert damping
and of the gyromagnetic ratio.
In section~\ref{sec_cit_formalism} we present the formalism
used to calculate the current-induced torques.
In sections~\ref{sec_results_gyro},~\ref{sec_results_damping}, 
and~\ref{sec_results_cit}  we discuss the gyromagnetic ratio,
the Gilbert damping, and the current-induced torques in the
one-dimensional Rashba model, respectively. 
This paper ends with a summary in section~\ref{sec_summary}.
\section{Formalism}
\label{sec_formalism}
\subsection{One-dimensional Rashba model}
\label{sec_rashba_onedim}
The two-dimensional Rashba model is
given by the Hamiltonian~\cite{rashba_review}
\bege\label{eq_rashba_model_twodim}
\begin{aligned}
H&=-\frac{\hbar^2}{2m_e}
\frac{\partial^2}{\partial x^2}
-\frac{\hbar^2}{2m_e}
\frac{\partial^2}{\partial y^2}+\\
&+i
\alpha^{\rm R}
\sigma_{y}\frac{\partial}{\partial x}
-i
\alpha^{\rm R}
\sigma_{x}\frac{\partial}{\partial y}
+
\frac{\Delta V}{2}
\vn{\sigma}
\cdot
\magdir(\vn{r}),
\end{aligned}
\ee
where the first line describes the kinetic energy,
the first two terms in the second line describe the Rashba SOI and the
last term in the second line describes the exchange
splitting. $\magdir(\vn{r})$ is the
magnetization direction, which may depend on the
position $\vn{r}=(x,y)$, and $\vn{\sigma}$ is the vector of
Pauli spin matrices. 
By removing the terms 
with the $y$-derivatives 
from Eq.~\eqref{eq_rashba_model_twodim}, 
i.e., $-\frac{\hbar^2}{2m_e}\frac{\partial^2}{\partial y^2}$ 
and $-i\alpha^{\rm R}\sigma_{x}\frac{\partial}{\partial y}$,
one obtains a one-dimensional variant of the Rashba model with
the Hamiltonian~\cite{geotexmosoi}
\bege\label{eq_rashba_model_onedim}
H=-\frac{\hbar^2}{2m_e}
\frac{\partial^2}{\partial x^2}
+i
\alpha^{\rm R}
\sigma_{y}\frac{\partial}{\partial x}
+
\frac{\Delta V}{2}
\vn{\sigma}
\cdot
\magdir(x).
\ee

Eq.~\eqref{eq_rashba_model_onedim}
is invariant under the simultaneous rotation of $\vn{\sigma}$ and of
the magnetization $\magdir$ around the $y$ axis.
Therefore, if $\magdir(x)$ describes a flat cycloidal spin-spiral
propagating into the $x$ direction, as given by 
\bege\label{eq_spin_spiral_cycloid}
\hat{\vn{M}}(x)=
\begin{pmatrix}
\sin(qx)\\
0\\
\cos(qx)
\end{pmatrix},\\
\ee
we can use the unitary transformation
\bege\label{eq_gauge_trafo_matrix}
\mathcal{U}(x)=
\left(
\begin{array}{cc}
\cos(\frac{qx}{2}) &-\sin(\frac{qx}{2})\\[6pt]
\sin(\frac{qx}{2}) &\cos(\frac{qx}{2})
\end{array}
\right)
\ee
in order to transform Eq.~\eqref{eq_rashba_model_onedim}
into a position-independent effective Hamiltonian~\cite{geotexmosoi}:
\bege\label{eq_gauge_transformed_hamil}
H=\frac{1}{2m}
\left(
p_x+eA^{\rm eff}_{x}
\right)^2-\frac{m (\alpha^{\rm R})^2}{2  \hbar^2}
+\frac{\Delta V}{2}\sigma_{z}
,
\ee
where $p_x=-i\hbar \partial/\partial x$ is the $x$ component of the
momentum operator and
\bege\label{eq_effective_vector_potential}
A^{\rm eff}_{x}=-\frac{m }{e  \hbar}
\left(
\alpha^{\rm R}+\frac{\hbar^2}{2m}q
\right)
\sigma_{y}
\ee
is the $x$-component of the effective magnetic vector potential.
Eq.~\eqref{eq_effective_vector_potential} 
shows that the noncollinearity described by $q$ acts like an effective
SOI in the special case of the one-dimensional Rashba model.
This
suggests to
introduce the concept of effective SOI strength
\bege\label{eq_effective_soi}
\alpha_{\rm eff}^{\rm R}=
\alpha^{\rm R}_{\phantom{e}}+\frac{\hbar^2}{2m}q.
\ee
Based on this concept of the effective SOI strength one can obtain the
$q$-dependence of the one-dimensional Rashba model from its
$\alpha^{\rm R}$-dependence at $q=0$.  
That a noncollinear magnetic texture provides
a nonrelativistic effective SOI has been found also in the
context of the intrinsic contribution to the
nonadiabatic torque in the absence of relativistic SOI, which
can be interpreted as a spin-orbit torque arising
from this effective SOI~\cite{intrinsic_spin_torque_without_soc}.
While
the Hamiltonian in Eq.~\eqref{eq_rashba_model_onedim} 
depends on position $x$ through
the position-dependence of the magnetization $\hat{\vn{M}}(x)$ in 
Eq.~\eqref{eq_spin_spiral_cycloid}, the effective Hamiltonian 
in Eq.~\eqref{eq_gauge_transformed_hamil} is not dependent
on $x$ and therefore easy to diagonalize.

\subsection{Gilbert damping and gyromagnetic ratio}
\label{sec_gilbert_and_gyromagnetic_formalism}
In collinear magnets damping and gyromagnetic ratio can be
extracted from the tensor~\cite{invsot}
\bege\label{eq_lambda_torque}
\Lambda_{ij}=-
\frac{1}{V}\!
\lim_{\omega\to 0}\!
\frac{{\rm Im}G_{\mathcal{T}_{i},\mathcal{T}_{j}}^{\rm R}
(\hbar\omega)}{\hbar\omega},
\ee
where $V$ is the volume of the unit cell and
\bege\label{eq_torque_torque_correlation}
G_{
\mathcal{T}_{i},
\mathcal{T}_{j}
}^{\rm R}(\hbar\omega)=
-i\int\limits_{0}^{\infty}\rmd t e^{i\omega t}
\left\langle
[
\mathcal{T}_{i}(t),
\mathcal{T}_{j}(0)
]_{-}
\right\rangle
\ee
is the retarded torque-torque correlation function.
$\mathcal{T}_{i}$ is the $i$-th component of the 
torque operator~\cite{invsot}.
The dc-limit $\omega\rightarrow 0$ in Eq.~\eqref{eq_lambda_torque}
is only justified when the frequency of the magnetization dynamics, e.g., the
ferromagnetic resonance frequency, is smaller than the relaxation rate of the
electronic states. In thin magnetic layers and monoatomic chains on substrates
this is typically the case due to the strong interfacial disorder.  
However, in very pure crystalline samples at low temperatures the relaxation
rate may be smaller than the ferromagnetic resonance frequency and one
needs to assume $\omega>0$ in 
Eq.~\eqref{eq_lambda_torque}~\cite{absence_intraband_gilbert_damping,breakdown_adiabatic_approach_damping}.  
The tensor $\vn{\Lambda}$ depends on the 
magnetization direction $\hat{\vn{M}}$ and we 
decompose it into the tensor $\vn{S}$, 
which is even under magnetization
reversal ($\vn{S}(\hat{\vn{M}})=\vn{S}(-\hat{\vn{M}})$), 
and the tensor $\vn{A}$, 
which is odd under magnetization 
reversal ($\vn{A}(\hat{\vn{M}})=-\vn{A}(-\hat{\vn{M}})$),
such that $\vn{\Lambda}=\vn{S}+\vn{A}$,
where
\bege
S_{ij}(\hat{\vn{M}})=\frac{1}{2}
\left[
\Lambda_{ij}(\hat{\vn{M}})
+
\Lambda_{ij}(-\hat{\vn{M}})
\right]
\ee
and
\bege
A_{ij}(\hat{\vn{M}})=\frac{1}{2}
\left[
\Lambda_{ij}(\hat{\vn{M}})
-
\Lambda_{ij}(-\hat{\vn{M}})
\right].
\ee
One can show that $\vn{S}$ is 
symmetric, i.e., $S_{ij}(\hat{\vn{M}})=S_{ji}(\hat{\vn{M}})$,
while $\vn{A}$ is 
antisymmetric, i.e., $A_{ij}(\hat{\vn{M}})=-A_{ji}(\hat{\vn{M}})$.

The Gilbert damping may be extracted from the 
symmetric component $S$ 
as follows~\cite{invsot}:
\bege\label{eq_gilbert_from_lambda}
\alpha^{\rm G}_{ij}=\frac{|\gamma| S^{\phantom{e}}_{ij}}
{M\mu_{0}},
\ee
where $M$ is the magnetization. 
The gyromagnetic ratio $\gamma$ is obtained from $\Lambda$
according to the equation~\cite{invsot}
\bege\label{eq_gamma_from_lambda}
\frac{1}{\gamma}=\frac{1}{2\mu_0 M}
\sum_{ijk}\epsilon_{ijk}\Lambda_{ij}\hat{M}_{k}
=\frac{1}{2\mu_0 M}
\sum_{ijk}\epsilon_{ijk}A_{ij}\hat{M}_{k}.
\ee
It is convenient to discuss the gyromagnetic ratio in terms of
the dimensionless $g$-factor, which is related to $\gamma$ through
$\gamma=g\mu_0 \mu_{\rm B}/\hbar$.
Consequently, the
$g$-factor is given by
\bege\label{eq_gfac_from_lambda}
\frac{1}{g}=\frac{\mu_{\rm B}}{2\hbar M}
\sum_{ijk}\epsilon_{ijk}\Lambda_{ij}\hat{M}_{k}
=\frac{\mu_{\rm B}}{2\hbar M}
\sum_{ijk}\epsilon_{ijk}A_{ij}\hat{M}_{k}
.
\ee
Due to the presence of the Levi-Civita 
tensor $\epsilon_{ijk}$ 
in Eq.~\eqref{eq_gamma_from_lambda} 
and in Eq.~\eqref{eq_gfac_from_lambda}
the gyromagnetic ratio and
the $g$-factor
are determined solely by 
the antisymmetric 
component $\vn{A}$ 
of $\vn{\Lambda}$.

Various different conventions are used in the
literature concerning the sign of 
the $g$-factor~\cite{remarks_signs_gfactors}.
Here, we define the sign of the $g$-factor such that
$\gamma>0$ for $g>0$ and $\gamma<0$ for $g<0$.
According to Eq.~\eqref{eq_define_gamma} the
rate of change of the magnetic moment is therefore parallel 
to the torque for positive $g$ and antiparallel to the
torque for negative $g$. While we are interested in this work
in the spectroscopic $g$-factor, and hence in the
relation between the rate of change of the magnetic
moment and the torque, 
Ref.~\cite{remarks_signs_gfactors} discusses 
the relation between 
the magnetic moment $\vn{m}$ and the 
angular momentum $\vn{L}$ that
generates it, i.e., $\vn{m}=\gamma_{\rm static}\vn{L}$.
Since differentiation with respect to time and 
use of $\vn{T}=\rmd\vn{L}/\rmd t$
leads to Eq.~\eqref{eq_define_gamma} our definition
of the signs of $g$ and $\gamma$ agrees essentially with
the one suggested in Ref.~\cite{remarks_signs_gfactors},
which proposes to use a positive $g$ when the magnetic moment
is parallel to the angular momentum generating it and a negative $g$
when the magnetic moment is antiparallel to the angular momentum 
generating it.  

Combining Eq.~\eqref{eq_gilbert_from_lambda} 
and Eq.~\eqref{eq_gamma_from_lambda} we
can express the Gilbert damping in terms of 
$\vn{A}$ and $\vn{S}$ as follows:
\bege
\alpha^{\rm G}_{xx}=\frac{S_{xx}}{|A_{xy}|}.
\ee

In the independent particle approximation Eq.~\eqref{eq_lambda_torque} can be
written as ${\Lambda}^{\phantom{I}}_{ij}=
{\Lambda}^{\rm I(a)}_{ij}
+
{\Lambda}^{\rm I(b)}_{ij}
+
{\Lambda}^{\rm II}_{ij}$, where
\begin{gather}\label{eq_kubo_linear_response_totocorr}
\begin{aligned}
{\Lambda}^{\rm I(a)\phantom{I}}_{ij}\!\!\!\!&=
\phantom{-}\frac{1}{h}\intkspa
\,{\rm Tr}
\left\langle
\mathcal{T}_{i}
G^{\rm R}_{\vn{k}}(\mathcal{E}_{\rm F})
\mathcal{T}_{j}
G^{\rm A}_{\vn{k}}(\mathcal{E}_{\rm F})
\right\rangle
\\
{\Lambda}^{\rm I(b)\phantom{I}}_{ij}\!\!\!\!&=
-\frac{1}{h}\intkspa
\,{\rm Re}
\,{\rm Tr}
\left\langle
\mathcal{T}_{i}
G^{\rm R}_{\vn{k}}(\mathcal{E}_{\rm F})
\mathcal{T}_{j}
G^{\rm R}_{\vn{k}}(\mathcal{E}_{\rm F})
\right\rangle
\\
{\Lambda}^{\rm II\phantom{(a)}}_{ij}\!\!\!\!&=
\phantom{-}\frac{1}{h}
\intkspa
\int_{-\infty}^{\mathcal{E}_{\rm F}}
d\mathcal{E}
\,{\rm Re}
\,{\rm Tr}
\left\langle
\mathcal{T}_{i}G^{\rm R}_{\vn{k}}(\mathcal{E})
\mathcal{T}_{j}
\frac{dG^{\rm R}_{\vn{k}}(\mathcal{E})}{d\mathcal{E}}\right.\\
 &\quad\quad\quad\quad\quad\quad\quad\quad\,-\left.
\mathcal{T}_{i}\frac{dG^{\rm R}_{\vn{k}}(\mathcal{E})}{d\mathcal{E}}
\mathcal{T}_{j}
G^{\rm R}_{\vn{k}}(\mathcal{E})
\right\rangle.
\end{aligned}\raisetag{5.6\baselineskip}
\end{gather}
Here, $d$ is the dimension ($d=1$ or $d=2$ or $d=3$), $G^{\rm R}_{\vn{k}}(\mathcal{E})$ is the
retarded Green's function 
and $G^{\rm A}_{\vn{k}}(\mathcal{E})=[G^{\rm R}_{\vn{k}}(\mathcal{E})]^{\dagger}$.
$\mathcal{E}_{\rm F}$ is the Fermi energy.
${\Lambda}^{\rm I(b)\phantom{I}}_{ij}$ is symmetric under the
interchange of the indices $i$ and $j$
while ${\Lambda}^{\rm II}_{ij}$ is antisymmetric. 
The term ${\Lambda}^{\rm I(a)\phantom{I}}_{ij}$
contains both symmetric and antisymmetric components.
Since the Gilbert damping tensor is symmetric,
both ${\Lambda}^{\rm I(b)\phantom{I}}_{ij}$
and ${\Lambda}^{\rm I(a)\phantom{I}}_{ij}$
contribute to it.
Since the gyromagnetic tensor is antisymmetric,
both ${\Lambda}^{\rm II}_{ij}$
and ${\Lambda}^{\rm I(a)\phantom{I}}_{ij}$
contribute to it.

In order to account for disorder  
we use the Gaussian scalar disorder model, where the
scattering potential $\mathcal{V}(\vn{r})$ satisfies $\langle \mathcal{V}(\vn{r}) \rangle=0$ 
and $\langle \mathcal{V}(\vn{r}) \mathcal{V}(\vn{r}') \rangle=U\delta(\vn{r}-\vn{r}')$.
This model is frequently used to calculate transport properties
in disordered multiband model systems~\cite{sj_sinova},
but it has also been combined with \textit{ab-initio} electronic
structure calculations
to study the anomalous Hall effect~\cite{sj_juergen,ahe_philippe}
and the anomalous Nernst effect~\cite{sj_ane_juergen}
in transition metals and their alloys.

In the clean limit, i.e., in the
limit $U\rightarrow 0$, the 
antisymmetric contribution to Eq.~\eqref{eq_kubo_linear_response_totocorr}
can be written as $A^{\phantom{i}}_{ij}=A^{\rm int}_{ij}+A^{\rm scatt}_{ij}$,
where the intrinsic part is given by
\bege\label{eq_toto_clean_limit}
\begin{aligned}
A^{\rm int}_{ij}&=\hbar
\intkspa
\sum_{n,m}
[
f_{\vn{k}n}-f_{\vn{k}m}
]
{\rm Im}
\frac{
\mathcal{T}^{i}_{\vn{k}nm}
\mathcal{T}^{j}_{\vn{k}mn}
}
{(\mathcal{E}_{\vn{k}n}-\mathcal{E}_{\vn{k}m})^2}\\
&=2\hbar
\intkspa
\sum_{n}
\sum_{ll'}
f_{\vn{k}n}
{\rm Im}
\left[
\frac{
\partial
\langle u_{\vn{k}n}|
}{\partial \hat{M}_l}
\frac{\partial
|u_{\vn{k}n}\rangle
}{\partial \hat{M}_{l'}}
\right]\times\\
&\quad\quad\times
\sum_{mm'}
\epsilon_{ilm}
\epsilon_{jl'm'}
\hat{M}_m
\hat{M}_{m'}
.
\end{aligned}
\ee
The second line in Eq.~\eqref{eq_toto_clean_limit}
expresses $A^{\rm int}_{ij}$ in terms of the Berry curvature
in magnetization space~\cite{spin_dynamics_qian_vignale}.
The scattering contribution is given by
\bege\label{eq_toto_clean_limit_scatt}
\begin{aligned}
&A^{\rm scatt}_{ij}=
\hbar
\sum_{nm}
\intkspa
\delta(\mathcal{E}_{\rm F}-\mathcal{E}_{\vn{k}n})
{\rm Im}\Biggl\{\\
-
&\left[
\mathcal{M}^{i}_{\vn{k}nm}
\frac{\gamma^{\phantom{i}}_{\vn{k}mn}}{\gamma^{\phantom{i}}_{\vn{k}nn}}
\mathcal{T}^{j}_{\vn{k}nn}-
\mathcal{M}^{j}_{\vn{k}nm}
\frac{\gamma^{\phantom{i}}_{\vn{k}mn}}{\gamma^{\phantom{i}}_{\vn{k}nn}}
\mathcal{T}^{i}_{\vn{k}nn}
\right]\\
+
&\left[
\mathcal{M}^{i}_{\vn{k}mn}\tilde{\mathcal{T}}^{j}_{\vn{k}nm}
-
\mathcal{M}^{j}_{\vn{k}mn}\tilde{\mathcal{T}}^{i}_{\vn{k}nm}
\right]
\\
-
&\left[
\mathcal{M}^{i}_{\vn{k}nm}
\frac{\gamma^{\phantom{i}}_{\vn{k}mn}}{\gamma^{\phantom{i}}_{\vn{k}nn}}
\tilde{\mathcal{T}}^{j}_{\vn{k}nn}
-
\mathcal{M}^{j}_{\vn{k}nm}
\frac{\gamma^{\phantom{i}}_{\vn{k}mn}}{\gamma^{\phantom{i}}_{\vn{k}nn}}
\tilde{\mathcal{T}}^{i}_{\vn{k}nn}
\right]
\\
+
&\left[
\tilde{\mathcal{T}}^{i}_{\vn{k}nn}
\frac{\gamma^{\phantom{i}}_{\vn{k}nm}}{\gamma^{\phantom{i}}_{\vn{k}nn}}
\frac{
\tilde{\mathcal{T}}^{j}_{\vn{k}mn}
}{
\mathcal{E}_{\vn{k}n}-
\mathcal{E}_{\vn{k}m}
}
-
\tilde{\mathcal{T}}^{j}_{\vn{k}nn}
\frac{\gamma^{\phantom{i}}_{\vn{k}nm}}{\gamma^{\phantom{i}}_{\vn{k}nn}}
\frac{
\tilde{\mathcal{T}}^{i}_{\vn{k}mn}
}{
\mathcal{E}_{\vn{k}n}-
\mathcal{E}_{\vn{k}m}
}
\right]
\\
+\frac{1}{2}
&\left[
\tilde{\mathcal{T}}^{i}_{\vn{k}nm}
\frac{1}{
\mathcal{E}_{\vn{k}n}-
\mathcal{E}_{\vn{k}m}
}
\tilde{\mathcal{T}}^{j}_{\vn{k}mn}
-
\tilde{\mathcal{T}}^{j}_{\vn{k}nm}
\frac{1}{
\mathcal{E}_{\vn{k}n}-
\mathcal{E}_{\vn{k}m}
}
\tilde{\mathcal{T}}^{i}_{\vn{k}mn}
\right]\\
+
&\Bigl[
\mathcal{T}^{j}_{\vn{k}nn}
\frac{\gamma^{\phantom{i}}_{\vn{k}nm}}{\gamma^{\phantom{i}}_{\vn{k}nn}}
\frac{1}{
\mathcal{E}_{\vn{k}n}-
\mathcal{E}_{\vn{k}m}
}
\tilde{\mathcal{T}}^{i}_{\vn{k}mn}\\
\quad\quad&-
\mathcal{T}^{i}_{\vn{k}nn}
\frac{\gamma^{\phantom{i}}_{\vn{k}nm}}{\gamma^{\phantom{i}}_{\vn{k}nn}}
\frac{1}{
\mathcal{E}_{\vn{k}n}-
\mathcal{E}_{\vn{k}m}
}
\tilde{\mathcal{T}}^{j}_{\vn{k}mn}
\Bigr]
\Biggr\}.\\
\end{aligned}
\ee
Here,
$\mathcal{T}^{i}_{\vn{k}nm}=
\langle
u_{\vn{k}n}
|
\mathcal{T}_{i}{\phantom{i}}
|
u_{\vn{k}m}
\rangle
$
are the matrix elements of the torque operator.
 $\tilde{\mathcal{T}}^{i}_{\vn{k}nm}$
denotes the
vertex corrections 
of the torque, which 
solve the equation
\bege
\begin{aligned}
\tilde{\mathcal{T}}^{i}_{\vn{k}nm}=&
\sum_{p}
\int\frac{\rmd^{n}k'}{(2\pi)^{n-1}}
\frac{
\delta(
\mathcal{E}_{F}-\mathcal{E}_{\vn{k}'p}
)
}
{2\gamma^{\phantom{i}}_{\vn{k}'pp}}\times
\\
\times
&
\langle
u_{\vn{k}n}
|
u_{\vn{k}'p}
\rangle
\left[
\tilde{\mathcal{T}}^{i}_{\vn{k}'pp}
+
\mathcal{T}^{i}_{\vn{k}'pp}
\right]
\langle
u_{\vn{k}'p}
|
u_{\vn{k}m}
\rangle.
\\
\end{aligned}
\ee
The matrix $\gamma^{\phantom{i}}_{\vn{k}nm}$ is given by
\bege
\gamma^{\phantom{i}}_{\vn{k}nm}=-\pi\sum_{p}
\intkspapri\delta(\mathcal{E}_{\rm F}-\mathcal{E}_{\vn{k}'p})
\langle
u_{\vn{k}n}
|
u_{\vn{k}'p}
\rangle
\langle
u_{\vn{k}'p}
|
u_{\vn{k}m}
\rangle
\ee
and the Berry connection in magnetization space is defined as
\bege
i\mathcal{M}_{\vn{k}nm}^{j}=i\frac{
\mathcal{T}_{\vn{k}nm}^{j}
}
{
\mathcal{E}_{\vn{k}m}-
\mathcal{E}_{\vn{k}n}
}.
\ee
The scattering contribution
Eq.~\eqref{eq_toto_clean_limit_scatt} formally resembles the
side-jump contribution to the AHE~\cite{sj_sinova} as obtained
from the scalar disorder model: 
It can be obtained by replacing the velocity operators
in Ref.~\cite{sj_sinova} by torque operators.
We find that in collinear magnets without SOI 
this scattering contribution 
vanishes. The gyromagnetic ratio is then given purely by the
intrinsic contribution Eq.~\eqref{eq_toto_clean_limit}. 
This is an interesting difference to the AHE:
Without SOI all contributions to the AHE are zero in collinear magnets, 
while both the intrinsic and the side-jump contributions
are generally nonzero in the presence of SOI. 

In the absence of SOI
Eq.~\eqref{eq_toto_clean_limit} can be expressed in terms of the
magnetization~\cite{spin_dynamics_qian_vignale}:
\bege\label{eq_a_intrinsic}
A^{\rm int}_{ij}
=-\frac{\hbar}{2\mu_{\rm B}}\sum_{k}
\epsilon_{ijk}
M_{k}
.
\ee
Inserting Eq.~\eqref{eq_a_intrinsic} into Eq.~\eqref{eq_gfac_from_lambda} yields 
$g=-2$, i.e., the expected
nonrelativistic value of the $g$-factor.

The $g$-factor in the presence of SOI is usually assumed 
to be given by~\cite{gyromagnetic_ratio_kittel}
\bege\label{eq_model_gfactor}
g=-2\frac{M_{\rm spin}+M_{\rm orb}}{M_{\rm spin}}
=-2\frac{M}{M_{\rm spin}}
,
\ee 
where $M_{\rm orb}$ is the orbital magnetization, $M_{\rm spin}$ is the
spin magnetization and $M=M_{\rm orb}+M_{\rm spin}$ is the total
magnetization. The
$g$-factor obtained from
Eq.~\eqref{eq_model_gfactor} 
is usually in good agreement with experimental 
results~\cite{magnetic_properties_ultrathin_3d_transition_metal_binary_alloys}. 
When SOI is absent, the orbital magnetization is zero, $M_{\rm orb}=0$, and
consequently Eq.~\eqref{eq_model_gfactor} yields $g=-2$ in that case.
Eq.~\eqref{eq_gfac_from_lambda} can be rewritten as
\bege\label{eq_gfac_from_lambda_version2}
\begin{aligned}
\frac{1}{g}
&=
\frac{M_{\rm spin}}{M}
\frac{\mubo}{2\hbar M_{\rm spin}}
\sum_{ijk}\epsilon_{ijk}A_{ij}\hat{M}_{k}=
\frac{M_{\rm spin}}{M}
\frac{1}{g_1^{\phantom{1}}},
\end{aligned}
\ee
with
\bege\label{eq_gfactor1}
\frac{1}{g_1^{\phantom{1}}}=\frac{\mubo}{2\hbar M_{\rm spin}}
\sum_{ijk}\epsilon_{ijk}A_{ij}\hat{M}_{k}.
\ee
From the comparison of Eq.~\eqref{eq_gfac_from_lambda_version2}
with Eq.~\eqref{eq_model_gfactor} it follows that Eq.~\eqref{eq_model_gfactor}
holds exactly if $g_1^{\phantom{1}}=-2$ is satisfied. However, 
Eq.~\eqref{eq_gfactor1} usually yields $g_1^{\phantom{1}}=-2$ only 
in collinear magnets when SOI is
absent, otherwise $g_1^{\phantom{1}}\ne -2$.
In the one-dimensional Rashba model the orbital magnetization 
is zero, $M_{\rm orb}=0$, and consequently
\bege
\frac{1}{g}=\frac{\mubo}{2\hbar M_{\rm spin}}
\sum_{ijk}\epsilon_{ijk}A_{ij}\hat{M}_{k}.
\ee

The symmetric contribution can be written 
as $S^{\phantom{i}}_{ij}=S^{\rm int}_{ij}+S^{\rm RR-vert}_{ij}+S^{\rm RA-vert}_{ij}$, where
\bege\label{eq_s_int}
\begin{aligned}
S^{\rm int}_{ij}=\frac{1}{h}
\intkspa
{\rm Tr}\left\{
\mathcal{T}_{i}
G_{\vn{k}}^{\rm R}(\mathcal{E}_{\rm F})
\mathcal{T}_{j}
\left[
G_{\vn{k}}^{\rm A}(\mathcal{E}_{\rm F})
-
G_{\vn{k}}^{\rm R}(\mathcal{E}_{\rm F})
\right]
\right\}
\end{aligned}
\ee
and
\bege\label{eq_s_rr_vert}
\begin{aligned}
S^{\rm RR-vert}_{ij}=-\frac{1}{h}
\intkspa
{\rm Tr}\left\{
\tilde{\mathcal{T}}^{\rm RR}_{i}
G_{\vn{k}}^{\rm R}(\mathcal{E}_{\rm F})
\mathcal{T}^{\phantom{i}}_{j}
G_{\vn{k}}^{\rm R}(\mathcal{E}_{\rm F})
\right\}
\end{aligned}
\ee
and
\bege\label{eq_s_ra_vert}
\begin{aligned}
S^{\rm AR-vert}_{ij}=\frac{1}{h}
\intkspa
{\rm Tr}\left\{
\tilde{\mathcal{T}}^{\rm AR}_{i}
G_{\vn{k}}^{\rm R}(\mathcal{E}_{\rm F})
\mathcal{T}^{\phantom{i}}_{j}
G_{\vn{k}}^{\rm A}(\mathcal{E}_{\rm F})
\right\},
\end{aligned}
\ee
where $G^{\rm R}_{\vn{k}}(\mathcal{E}_{\rm F})=\hbar[
\mathcal{E}^{\phantom{i}}_{\rm F}-H^{\phantom{i}}_{\vn{k}}-
\Sigma_{\vn{k}}^{\rm R}(\mathcal{E}_{\rm F})
]^{-1}$ is the retarded Green's function,
$G_{\vn{k}}^{\rm A}(\mathcal{E}_{\rm F})=
\left[G_{\vn{k}}^{\rm R}(\mathcal{E}_{\rm F})\right]^{\dagger}
$
is the advanced Green's function
and
\bege
\Sigma^{\rm R}(\mathcal{E}_{\rm F})
=
\frac{U}{\hbar}
\intkspa
G_{\vn{k}}^{\rm R}(\mathcal{E}_{\rm F})
\ee
is the retarded self-energy.
The vertex corrections are determined by the
equations
\bege\label{eq_vertex_corr_torque_ar}
\tilde{
\vn{\mathcal{T}}
}^{\rm AR}=
\vn{\mathcal{T}}
+
\frac{U}{\hbar^2}
\intkspa
G_{\vn{k}}^{\rm A}(\mathcal{E}_{\rm F})
\tilde{
\vn{\mathcal{T}}
}^{\rm AR}_{\vn{k}}
G_{\vn{k}}^{\rm R}(\mathcal{E}_{\rm F})
\ee
and
\bege\label{eq_vertex_corr_torque_rr}
\tilde{
\vn{\mathcal{T}}
}^{\rm RR}=
\vn{\mathcal{T}}
+\frac{U}{\hbar^2}
\intkspa
G_{\vn{k}}^{\rm R}(\mathcal{E}_{\rm F})
\tilde{
\vn{\mathcal{T}}
}^{\rm RR}_{\vn{k}}
G_{\vn{k}}^{\rm R}(\mathcal{E}_{\rm F}).
\ee

In contrast to the antisymmetric tensor $A$, which becomes independent
of the scattering strength $U$ for sufficiently small $U$, i.e.,
in the clean limit, the symmetric tensor $S$ depends strongly on $U$
in metallic systems
in the clean limit. $S^{\rm int}_{ij}$ and $S^{\rm scatt}_{ij}$ depend 
therefore on $U$ through the self-energy
and through the vertex corrections.

In the case of the one-dimensional Rashba model,
the equations Eq.~\eqref{eq_toto_clean_limit} and Eq.~\eqref{eq_toto_clean_limit_scatt}
for the antisymmetric tensor $A$ and the equations 
Eq.~\eqref{eq_s_int}, Eq.~\eqref{eq_s_rr_vert} and Eq.~\eqref{eq_s_ra_vert} for the
symmetric tensor $S$ can be used both for the collinear magnetic state as well as for 
the spin-spiral of Eq.~\eqref{eq_spin_spiral_cycloid}. 
To obtain the $g$-factor for the collinear magnetic state, we plug the eigenstates 
and eigenvalues of Eq.~\eqref{eq_rashba_model_onedim} 
(with $\hat{\vn{M}}=\hat{\vn{e}}_{z}$) into 
Eq.~\eqref{eq_toto_clean_limit} and into Eq.~\eqref{eq_toto_clean_limit_scatt}.
In the case of the spin-spiral of Eq.~\eqref{eq_spin_spiral_cycloid} we use instead
the eigenstates and eigenvalues of Eq.~\eqref{eq_gauge_transformed_hamil}.
Similarly, to obtain the Gilbert damping in the collinear magnetic state, we 
evaluate Eq.~\eqref{eq_s_int}, Eq.~\eqref{eq_s_rr_vert} and Eq.~\eqref{eq_s_ra_vert}
based on the Hamiltonian in Eq.~\eqref{eq_rashba_model_onedim} and
for the spin-spiral we use instead the effective Hamiltonian 
in Eq.~\eqref{eq_gauge_transformed_hamil}.

\subsection{Current-induced torques}
\label{sec_cit_formalism}
The current-induced torque on the magnetization
can be expressed in terms of the torkance tensor $t_{ij}$
as~\cite{ibcsoit}
\bege\label{eq_torque_and_torkance}
T_{i}=\sum_{j}t_{ij}E_{j},
\ee
where $E_{j}$ is the $j$-th component of the 
applied electric field and $T_{i}$ is the $i$-th component
of the torque per volume~\cite{note_on_units}. 
$t_{ij}$ is the sum of
three terms, ${t}^{\phantom{I}}_{ij}=
{t}^{\rm I(a)}_{ij}
+
{t}^{\rm I(b)}_{ij}
+
{t}^{\rm II}_{ij}$, where~\cite{ibcsoit}
\begin{gather}\label{eq_kubo_linear_response_tovelocorr}
\begin{aligned}
t^{\rm I(a)\phantom{I}}_{ij}\!\!\!\!&=
\phantom{-}\frac{e}{h}\intkspa
\,{\rm Tr}
\left\langle
\mathcal{T}_{i}
G^{\rm R}_{\vn{k}}(\mathcal{E}_{\rm F})
v_{j}
G^{\rm A}_{\vn{k}}(\mathcal{E}_{\rm F})
\right\rangle
\\
t^{\rm I(b)\phantom{I}}_{ij}\!\!\!\!&=
-\frac{e}{h}\intkspa
\,{\rm Re}
\,{\rm Tr}
\left\langle
\mathcal{T}_{i}
G^{\rm R}_{\vn{k}}(\mathcal{E}_{\rm F})
v_{j}
G^{\rm R}_{\vn{k}}(\mathcal{E}_{\rm F})
\right\rangle
\\
t^{\rm II\phantom{(a)}}_{ij}\!\!\!\!&=
\phantom{-}\frac{e}{h}
\intkspa
\int_{-\infty}^{\mathcal{E}_{\rm F}}
d\mathcal{E}
\,{\rm Re}
\,{\rm Tr}
\left\langle
\mathcal{T}_{i}G^{\rm R}_{\vn{k}}(\mathcal{E})
v_{j}
\frac{dG^{\rm R}_{\vn{k}}(\mathcal{E})}{d\mathcal{E}}\right.\\
 &\quad\quad\quad\quad\quad\quad\quad\quad\,-\left.
\mathcal{T}_{i}\frac{dG^{\rm R}_{\vn{k}}(\mathcal{E})}{d\mathcal{E}}
v_{j}
G^{\rm R}_{\vn{k}}(\mathcal{E})
\right\rangle.
\end{aligned}\raisetag{5.6\baselineskip}
\end{gather}
We decompose the torkance into two parts that are,
respectively, even and odd
with respect to magnetization 
reversal, i.e., $t_{ij}^{\rm e}(\hat{\vn{M}})=[
t^{\phantom{i}}_{ij}(\hat{\vn{M}})+
t^{\phantom{i}}_{ij}(-\hat{\vn{M}})]/2$
and $t_{ij}^{\rm o}(\hat{\vn{M}})=
[t^{\phantom{i}}_{ij}(\hat{\vn{M}})-
t^{\phantom{i}}_{ij}(-\hat{\vn{M}})]/2$.

In the clean limit, i.e., 
for $U\rightarrow 0$, the even torkance can be 
written as $t^{\rm e}_{ij}=t^{\rm e,int}_{ij}+t^{\rm e,scatt}_{ij}$,
where~\cite{ibcsoit}
\bege\label{eq_tovelo_clean_limit_intrinsic}
t^{\rm e,int}_{ij}=2e\hbar
\intkspa
\sum_{n\ne m}
f_{\vn{k}n}
{\rm Im}
\frac{
\mathcal{T}^{i}_{\vn{k}nm}
v^{j}_{\vn{k}mn}
}
{(\mathcal{E}_{\vn{k}n}-\mathcal{E}_{\vn{k}m})^2}
\ee
is the intrinsic contribution and
\bege\label{eq_tovelo_clean_limit_scatt}
\begin{aligned}
&t^{\rm e,scatt}_{ij}=e\hbar
\sum_{nm}
\intkspa
\delta(\mathcal{E}_{\rm F}-\mathcal{E}_{\vn{k}n})
{\rm Im}\Biggl\{\\
&\Bigl[
-\mathcal{M}^{i}_{\vn{k}nm}
\frac{\gamma^{\phantom{i}}_{\vn{k}mn}}
{\gamma^{\phantom{i}}_{\vn{k}nn}}
v_{\vn{k}nn}^{j}+
\mathcal{A}^{j}_{\vn{k}nm}
\frac{\gamma^{\phantom{i}}_{\vn{k}mn}}
{\gamma^{\phantom{i}}_{\vn{k}nn}}
\mathcal{T}_{\vn{k}nn}^{i}
\Bigr]\\
+&\Bigl[\mathcal{M}^{i}_{\vn{k}mn}\tilde{v}^{j}_{\vn{k}nm}
-
\mathcal{A}^{j}_{\vn{k}mn}\tilde{\mathcal{T}}^{i}_{\vn{k}nm}
\Bigr]\\
-
&\Bigl[
\mathcal{M}^{i}_{\vn{k}nm}
\frac{\gamma^{\phantom{i}}_{\vn{k}mn}}{\gamma^{\phantom{i}}_{\vn{k}nn}}
\tilde{v}^{j}_{\vn{k}nn}
-
\mathcal{A}^{j}_{\vn{k}nm}
\frac{\gamma^{\phantom{i}}_{\vn{k}mn}}{\gamma^{\phantom{i}}_{\vn{k}nn}}
\tilde{\mathcal{T}}^{i}_{\vn{k}nn}
\Bigr]\\
+
&\Bigl[\tilde{v}^{j}_{\vn{k}mn}
\frac{\gamma^{\phantom{i}}_{\vn{k}nm}}
{\gamma^{\phantom{i}}_{\vn{k}nn}}
\frac{
\tilde{\mathcal{T}}^{i}_{nn}
}{
\mathcal{E}_{\vn{k}n}-
\mathcal{E}_{\vn{k}m}
}
-
\tilde{\mathcal{T}}^{i}_{\vn{k}mn}
\frac{\gamma^{\phantom{i}}_{\vn{k}nm}}
{\gamma^{\phantom{i}}_{\vn{k}nn}}
\frac{
\tilde{v}^{j}_{\vn{k}nn}
}{
\mathcal{E}_{\vn{k}n}-
\mathcal{E}_{\vn{k}m}
}
\Bigr]\\
+\frac{1}{2}
&\Bigl[\tilde{v}^{j}_{\vn{k}nm}
\frac{1}{
\mathcal{E}_{\vn{k}n}-
\mathcal{E}_{\vn{k}m}
}
\tilde{\mathcal{T}}^{i}_{\vn{k}mn}
-
\tilde{\mathcal{T}}^{i}_{\vn{k}nm}
\frac{1}{
\mathcal{E}_{\vn{k}n}-
\mathcal{E}_{\vn{k}m}
}
\tilde{v}^{j}_{\vn{k}mn}
\Bigr]\\
+
&\Bigl[v^{j}_{\vn{k}nn}
\frac{\gamma^{\phantom{i}}_{\vn{k}nm}}
{\gamma^{\phantom{i}}_{\vn{k}nn}}
\frac{1}{
\mathcal{E}_{\vn{k}n}-
\mathcal{E}_{\vn{k}m}
}
\tilde{\mathcal{T}}^{i}_{\vn{k}mn}
\\
&\quad\quad
-\mathcal{T}^{i}_{\vn{k}nn}
\frac{\gamma^{\phantom{i}}_{\vn{k}nm}}
{\gamma^{\phantom{i}}_{\vn{k}nn}}
\frac{1}{
\mathcal{E}_{\vn{k}n}-
\mathcal{E}_{\vn{k}m}
}
\tilde{v}^{j}_{\vn{k}mn}
\Bigr]
\Biggr\}.\\
\end{aligned}
\ee
is the scattering contribution.
Here,
\bege
i\mathcal{A}_{\vn{k}nm}^{j}=i\frac{
v_{\vn{k}nm}^{j}
}
{
\mathcal{E}_{\vn{k}m}-
\mathcal{E}_{\vn{k}n}
}=
\frac{i}{\hbar}
\langle
u_{\vn{k}n}
|
\frac{
\partial
}
{
\partial k^j
}
|
u_{\vn{k}m}
\rangle
\ee
is the Berry connection in $\vn{k}$ space and
the vertex corrections of the velocity operator solve
the equation
\bege
\begin{aligned}
\tilde{v}^{i}_{\vn{k}nm}=&
\sum_{p}
\int\frac{\rmd^{n}k'}{(2\pi)^{n-1}}
\frac{
\delta(
\mathcal{E}_{F}-\mathcal{E}_{\vn{k}'p}
)
}
{2\gamma^{\phantom{i}}_{\vn{k}'pp}}
\times
\\
&
\times
\langle
u_{\vn{k}n}
|
u_{\vn{k}'p}
\rangle
\left[
\tilde{v}^{i}_{\vn{k}'pp}
+
v^{i}_{\vn{k}'pp}
\right]
\langle
u_{\vn{k}'p}
|
u_{\vn{k}m}
\rangle
.\\
\end{aligned}
\ee

The  odd contribution can be written 
as $t_{ij}^{\rm o}=t_{ij}^{\rm o,int}+t_{ij}^{\rm RR-vert}+t_{ij}^{\rm AR-vert}$,
where
\bege\label{eq_tovelo_odd_intrinsic}
t_{ij}^{\rm o,int}=\frac{e}{h}
\intkspa
{\rm Tr}\left\{
\mathcal{T}_{i}
G^{\rm R}_{\vn{k}}(\mathcal{E}_{\rm F})
v^{\phantom{i}}_{j}
\left[
G^{\rm A}_{\vn{k}}(\mathcal{E}_{\rm F})
-
G^{\rm R}_{\vn{k}}(\mathcal{E}_{\rm F})
\right]
\right\}
\ee
and
\bege\label{eq_tovelo_odd_vert_rr}
t_{ij}^{\rm RR-vert}=-\frac{e}{h}
\intkspa
{\rm Tr}\left\{
\tilde{\mathcal{T}}^{\rm RR}_{i}
G^{\rm R}_{\vn{k}}(\mathcal{E}_{\rm F})
v^{\phantom{i}}_{j}
G^{\rm R}_{\vn{k}}(\mathcal{E}_{\rm F})
\right\}
\ee
and
\bege\label{eq_tovelo_odd_vert_ra}
t_{ij}^{\rm AR-vert}=\frac{e}{h}
\intkspa
{\rm Tr}\left\{
\tilde{\mathcal{T}}^{\rm AR}_{i}
G^{\rm R}_{\vn{k}}(\mathcal{E}_{\rm F})
v^{\phantom{i}}_{j}
G^{\rm A}_{\vn{k}}(\mathcal{E}_{\rm F})
\right\}.
\ee
The vertex corrections $\tilde{\mathcal{T}}^{\rm AR}_{i}$ 
and $\tilde{\mathcal{T}}^{\rm RR}_{i}$ of the torque operator
are given in Eq.~\eqref{eq_vertex_corr_torque_ar} 
and in Eq.~\eqref{eq_vertex_corr_torque_rr}, respectively.

While the even torkance, Eq.~\eqref{eq_tovelo_clean_limit_intrinsic} 
and Eq.~\eqref{eq_tovelo_clean_limit_scatt}, becomes independent
of the scattering strength $U$ in the clean limit, i.e., 
for $U\rightarrow 0$, the odd torkance $t^{\rm o}_{ij}$
depends strongly on $U$ in metallic systems in the
clean limit~\cite{ibcsoit}.

In the case of the one-dimensional Rashba model,
the equations Eq.~\eqref{eq_tovelo_clean_limit_intrinsic} 
and Eq.~\eqref{eq_tovelo_clean_limit_scatt}
for the even torkance $t^{\rm e}_{ij}$ and the equations 
Eq.~\eqref{eq_tovelo_odd_intrinsic}, 
Eq.~\eqref{eq_tovelo_odd_vert_rr} and 
Eq.~\eqref{eq_tovelo_odd_vert_ra} for the
odd torkance $t^{\rm o}_{ij}$ can be used both for the collinear magnetic state as well as for 
the spin-spiral of Eq.~\eqref{eq_spin_spiral_cycloid}. 
To obtain the even torkance for the collinear magnetic state, we plug the eigenstates 
and eigenvalues of Eq.~\eqref{eq_rashba_model_onedim} 
(with $\hat{\vn{M}}=\hat{\vn{e}}_{z}$) into 
Eq.~\eqref{eq_tovelo_clean_limit_intrinsic} and 
into Eq.~\eqref{eq_tovelo_clean_limit_scatt}.
In the case of the spin-spiral of Eq.~\eqref{eq_spin_spiral_cycloid} we use instead
the eigenstates and eigenvalues of Eq.~\eqref{eq_gauge_transformed_hamil}.
Similarly, to obtain the odd torkance in the collinear magnetic state, we 
evaluate Eq.~\eqref{eq_tovelo_odd_intrinsic}, 
Eq.~\eqref{eq_tovelo_odd_vert_rr} and 
Eq.~\eqref{eq_tovelo_odd_vert_ra}
based on the Hamiltonian in Eq.~\eqref{eq_rashba_model_onedim} and
for the spin-spiral we use instead the effective Hamiltonian 
in Eq.~\eqref{eq_gauge_transformed_hamil}.

\section{Results}
\subsection{Gyromagnetic ratio}
\label{sec_results_gyro}
\begin{figure}
\includegraphics[width=\linewidth]{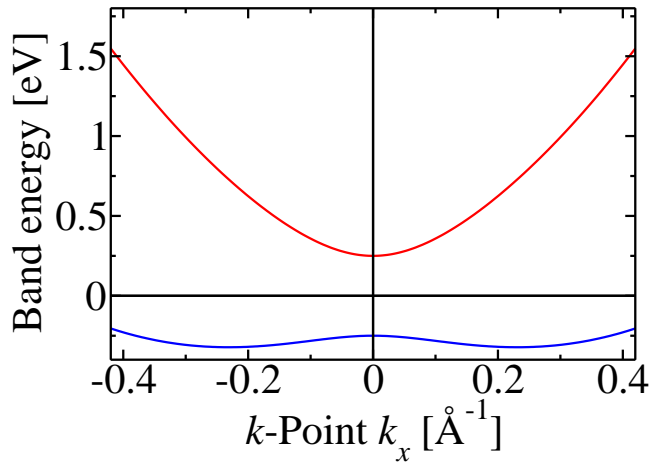}
\caption{\label{fig_bands_onedim}
Band structure of  
the one-dimensional Rashba model.
}
\end{figure}

\begin{figure}
\includegraphics[width=\linewidth]{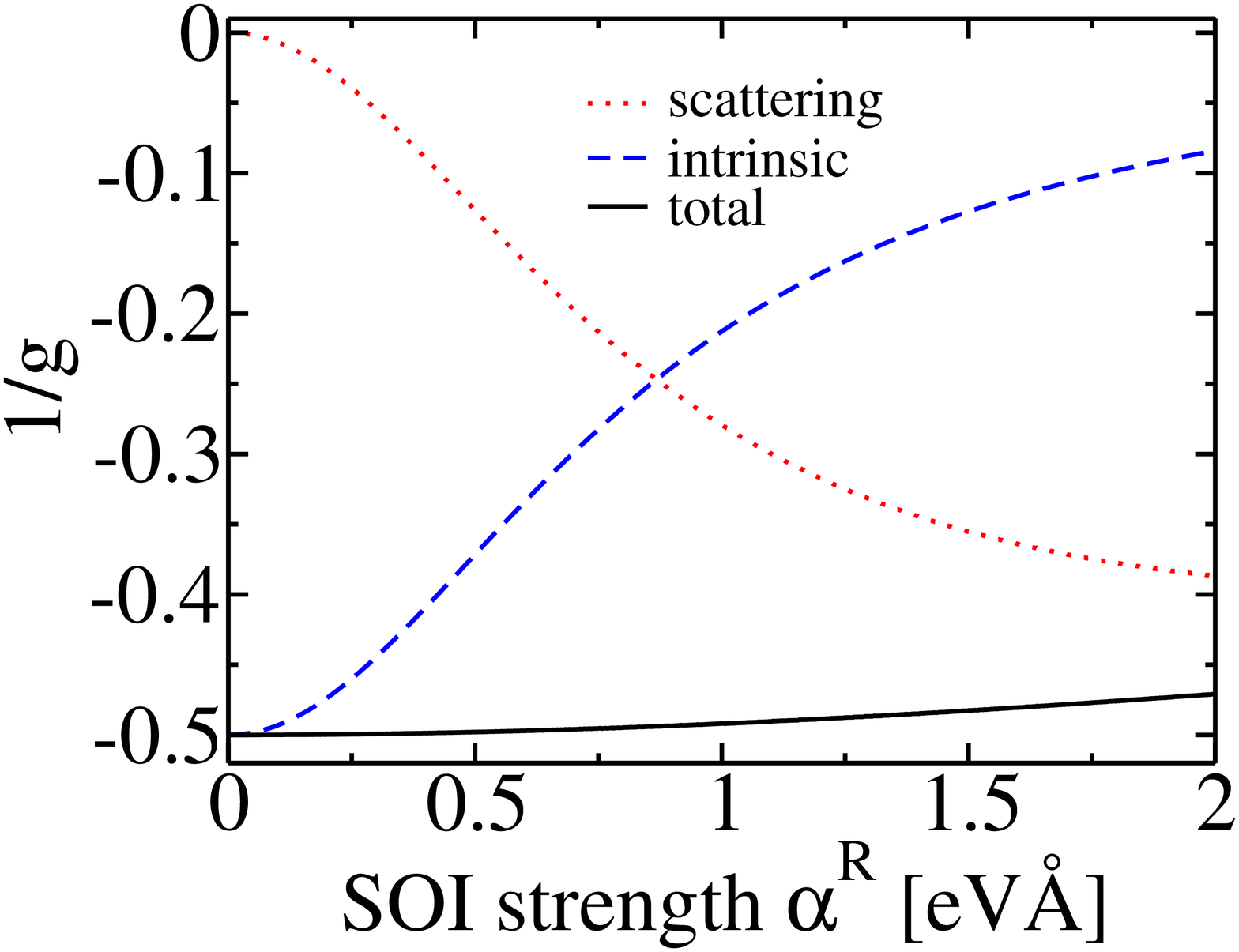}
\caption{\label{fig_onedim_rashba_collinear_vs_alpha}
Inverse $g$-factor vs.\ SOI strength $\alpha^{\rm R}$ in 
the one-dimensional Rashba model.
}
\end{figure}

We first discuss the $g$-factor in the collinear case, 
i.e., when $\magdir(\vn{r})=\hat{\vn{e}}_{z}$.
In this case the energy bands are given by
\bege\label{eq_energy_vs_k_rashba}
\mathcal{E}=\frac{\hbar^2 k^2_x}{2m}\pm\sqrt{\frac{1}{4}(\Delta V)^2+(\alpha^{\rm R}k_x)^2}.
\ee
When $\Delta V\ne 0$ or $\alpha^{\rm R}\ne 0$ the energy $\mathcal{E}$
can become negative.
The band structure of the one-dimensional Rashba
model is shown in Fig.~\ref{fig_bands_onedim} 
for the model parameters $\alpha^{\rm R}=$2eV\AA\,
and $\Delta V=0.5$eV.
For this choice of parameters the energy minima are
not located at $k_{x}=0$ but instead 
at 
\bege
k_{x}^{\rm min}=\pm
\frac{\sqrt{(\alpha^{\rm R})^4m^2-\frac{1}{4}\hbar^4(\Delta V)^2}}
{
\hbar^2
\alpha^{\rm R}
},
\ee
and the corresponding minimum of the energy is given by
\bege
\mathcal{E}^{\rm min}=
-\frac{
m(\alpha^{\rm R})^4
+
\frac{1}{4}
\frac{\hbar^4}{m}(\Delta V)^2
}
{
2\hbar^2
(\alpha^{\rm R})^2
}.
\ee  
The inverse $g$-factor is shown as a function of the  
SOI strength $\alpha^{\rm R}$ in
Fig.~\ref{fig_onedim_rashba_collinear_vs_alpha} for the exchange 
splitting $\Delta V=1$eV 
and Fermi energy $\mathcal{E}_{\rm F}=1.36$eV.
At $\alpha^{\rm R}=0$ the scattering contribution is zero, i.e., the
$g$-factor is determined completely by the intrinsic Berry 
curvature expression, Eq.~\eqref{eq_a_intrinsic}. 
Thus, at $\alpha^{\rm R}=0$ it assumes the value $1/g=-0.5$, which is the expected 
nonrelativistic value (see the discussion below Eq.~\eqref{eq_a_intrinsic}). 
With increasing SOI strength $\alpha^{\rm R}$ the intrinsic contribution 
to $1/g$ is more and more suppressed. 
However, the scattering contribution compensates this 
decrease such that the total $1/g$
is close to its nonrelativistic value of $-0.5$.
The neglect of the scattering corrections at large values
of $\alpha^{\rm R}$ would lead in this case to a
strong underestimation of  the magnitude of $1/g$, i.e., a strong
overestimation of the magnitude of $g$.
 
\begin{figure}
\includegraphics[width=\linewidth]{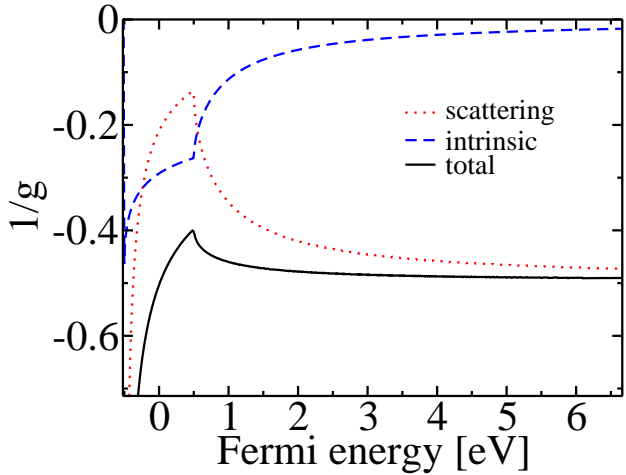}
\caption{\label{fig_onedim_rashba_collinear_vs_energy}
Inverse $g$-factor vs.\ Fermi energy in the one-dimensional Rashba model.
}
\end{figure}

However, at smaller values of the Fermi energy, the $g$ factor can
deviate substantially from its nonrelativistic value of $-2$.
To show this we plot in Fig.~\ref{fig_onedim_rashba_collinear_vs_energy}
the inverse $g$-factor
as a function of the Fermi energy when the exchange 
splitting 
and the SOI strength are set 
to $\Delta V=1$eV  
and $\alpha^{\rm R}=$2eV\AA, respectively.
As discussed in Eq.~\eqref{eq_energy_vs_k_rashba} the minimal
Fermi energy is negativ in this case.
The intrinsic contribution to $1/g$ declines with increasing Fermi
energy. At large values of the Fermi energy this decline is
compensated
by the increase of the vertex corrections and the total value of $1/g$
is close to $-0.5$.

\begin{figure}
\includegraphics[width=\linewidth]{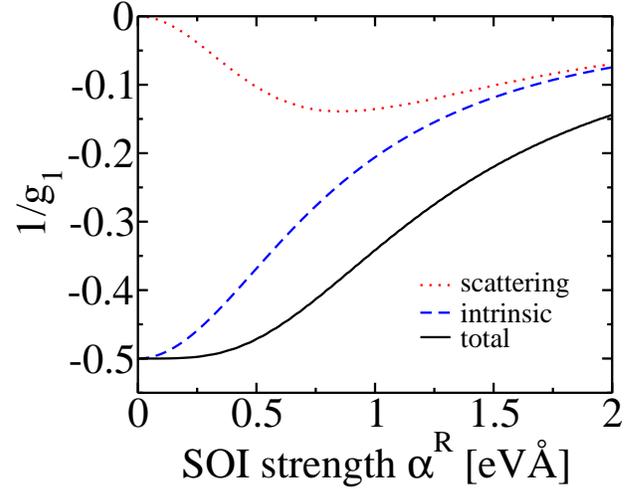}
\caption{\label{fig_twodim_rashba_collinear_vs_alpha}
Inverse $g_1$-factor vs.\ SOI strength $\alpha^{\rm R}$ in 
the two-dimensional Rashba model.
}
\end{figure}

Previous theoretical works on the $g$-factor 
have not considered the scattering 
contribution~\cite{transverse_dynamical_magnetic_susceptibility_g_shift}. 
It is therefore important to find out whether the compensation
of the decrease of the intrinsic contribution by the increase of the
extrinsic contribution as discussed in 
Fig.~\ref{fig_onedim_rashba_collinear_vs_alpha}
and Fig.~\ref{fig_onedim_rashba_collinear_vs_energy}
is peculiar to the one-dimensional Rashba model
or
whether it can be found in more general cases. 
For this reason we evaluate $g_1$ for
the
two-dimensional Rashba model. In Fig.~\ref{fig_twodim_rashba_collinear_vs_alpha}
we show the inverse $g_1$-factor in the two-dimensional Rashba model as
a function of SOI strength $\alpha^{\rm R}$ for the exchange splitting
$\Delta V=1$eV and the Fermi 
energy $\mathcal{E}_{\rm F}=1.36$eV. Indeed 
for $\alpha^{\rm R}<0.5$eV\AA\, 
the
scattering corrections tend to stabilize $g_1$ 
at its non-relativistic value.
However, in contrast to the one-dimensional 
case (Fig.~\ref{fig_onedim_rashba_collinear_vs_alpha}),
where $g$ does not deviate much from its
nonrelativistic value up to $\alpha^{\rm R}=2$eV\AA, $g_1$ 
starts to be affected by SOI at smaller 
values of  $\alpha^{\rm R}$ in the two-dimensional case.
According to Eq.~\eqref{eq_gfac_from_lambda_version2} 
the full $g$ factor is given by $g=g_{1}^{\phantom{1}}(1+M_{\rm orb}/M_{\rm spin})$.
Therefore, when the scattering corrections stabilize $g_{1}$ at its nonrelativistic value
the Eq.~\eqref{eq_model_gfactor} is satisfied.  In the two-dimensional
Rashba model $M_{\rm orb}=0$ when both bands are occupied. For the
Fermi energy $\mathcal{E}_{\rm F}=1.36$eV both bands are 
occupied and therefore $g=g_{1}$ for the range of parameters used
in Fig.~\ref{fig_twodim_rashba_collinear_vs_alpha}.

\begin{figure}
\includegraphics[width=\linewidth]{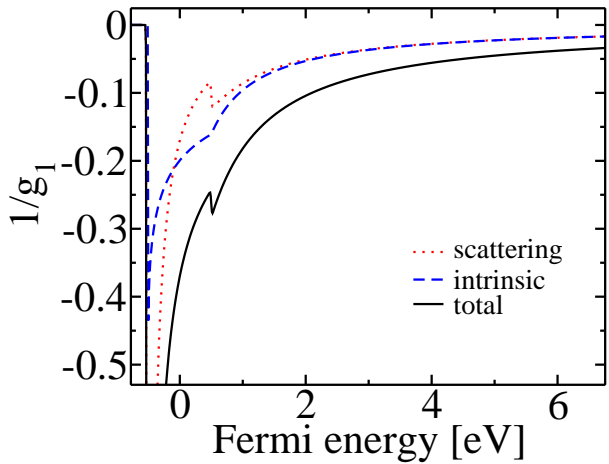}
\caption{\label{fig_twodim_rashba_gfactor}
Inverse $g_1$-factor $1/g_1$ vs.\ Fermi energy in 
the two-dimensional Rashba model.
}
\end{figure}

\begin{figure}
\includegraphics[width=\linewidth]{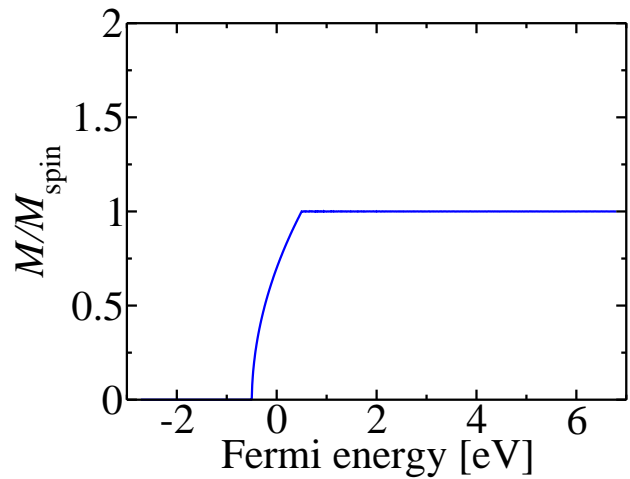}
\caption{\label{fig_magnetizationratio_vs_fermi}
Ratio of total magnetization and spin magnetization, $M/M_{\rm spin}$, 
vs.\ Fermi energy in 
the two-dimensional Rashba model.
}
\end{figure}

The inverse $g_1$ 
of the two-dimensional Rashba model 
is shown in Fig.~\ref{fig_twodim_rashba_gfactor} 
as a function of Fermi energy
for the
parameters $\Delta V=1$eV and $\alpha^{\rm R}=2$eV\AA.
The scattering correction is as large as the intrinsic contribution
when $\mathcal{E}_{\rm F}>1$eV. While the scattering correction
is therefore important, it is not sufficiently large to bring $g_1$
close to its nonrelativistic value in the energy range shown in
the figure, which is a major difference to the one-dimensional case
illustrated in Fig.~\ref{fig_onedim_rashba_collinear_vs_energy}.
According to Eq.~\eqref{eq_gfac_from_lambda_version2}
the $g$ factor is related to $g_1$ by $g=g_1 M/M_{\rm spin}$. 
Therefore, we show in Fig.~\ref{fig_magnetizationratio_vs_fermi}
the ratio $M/M_{\rm spin}$ as a function of Fermi energy.
At high Fermi energy (when both bands are occupied) 
the orbital magnetization is zero and $M/M_{\rm spin}=1$.
At low Fermi energy the sign of the orbital magnetization is opposite to
the sign of the spin magnetization such that the magnitude of $M$ is smaller
than the magnitude of $M_{\rm spin}$ resulting in the ratio $M/M_{\rm spin}<1$.

\begin{figure}
\includegraphics[width=\linewidth]{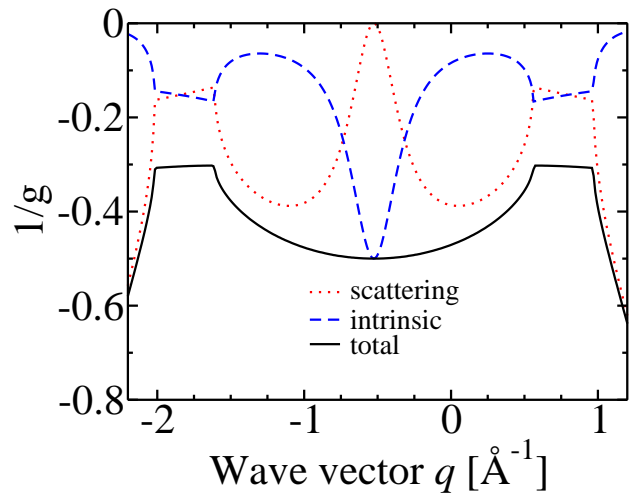}
\caption{\label{fig_gfac_onedim_rashba_vs_qvec_zerobroadening}
Inverse $g$-factor $1/g$ vs.\ wave number $q$ in 
the one-dimensional Rashba model. 
}
\end{figure}

Next, we discuss the $g$-factor of the 
one-dimensional Rashba model in the noncollinear case.
In Fig.~\ref{fig_gfac_onedim_rashba_vs_qvec_zerobroadening} we plot the inverse
$g$-factor and its decomposition into the intrinsic and scattering
contributions as a function of the spin-spiral wave vector $q$,
where exchange splitting, SOI strength and
Fermi energy are set 
to $\Delta V=1$eV, $\alpha^{\rm R}=2$eV\AA\,
and $\mathcal{E}_{\rm F}=1.36$eV, respectively. 
Since the curves are not symmetric around $q=0$,
the $g$-factor at wave number $q$ differs from the one at $-q$, i.e., 
the \textit{gyromagnetism in the Rashba model is chiral}. 
At $q=-2 m_e \alpha^{\rm R}/\hbar^2$
the $g$-factor assumes the
value of $g=-2$ and the scattering corrections are zero.
Moreover, the curves are symmetric 
around  $q=-2 m_e \alpha^{\rm R}/\hbar^2$.
These observations can be explained 
by the concept of the effective SOI
introduced in Eq.~\eqref{eq_effective_soi}:
At $q=-2 m_e \alpha^{\rm R}/\hbar^2$ the effective SOI is zero and
consequently the noncollinear magnet behaves like a collinear magnet
without SOI at this value of $q$. 
As we have discussed above 
in Fig.~\ref{fig_onedim_rashba_collinear_vs_alpha}, 
the $g$-factor of collinear magnets
is $g=-2$ when SOI
is absent, which explains why it is also $g=-2$ in
noncollinear magnets with
$q=-2 m_e \alpha^{\rm R}/\hbar^2$.
If only the intrinsic contribution is considered and
the scattering corrections are neglected, $1/g$ varies much
stronger around the point of zero 
effective SOI $q=-2 m_e \alpha^{\rm R}/\hbar^2$, i.e., the scattering
corrections stabilize $g$ at its nonrelativistic value close to the point
of zero effective SOI.

\subsection{Damping}
\label{sec_results_damping}
\begin{figure}
\includegraphics[width=\linewidth]{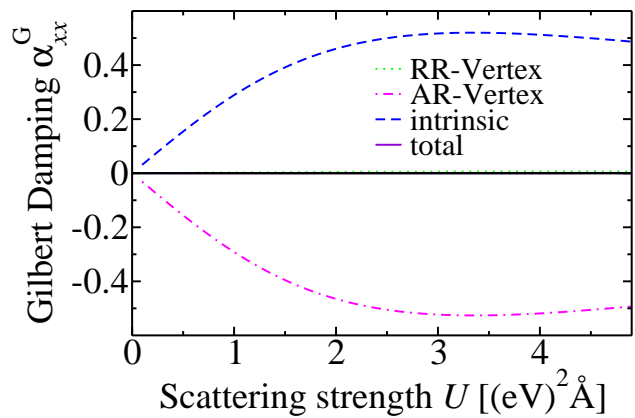}
\caption{\label{fig_damping_onedim_rashba_collinear_vs_scattering_nosoi}
Gilbert damping $\alpha^{\rm G}_{xx}$ vs.\ 
scattering strength $U$ 
in the one-dimensional Rashba model without SOI. In this case the
vertex corrections and the intrinsic contribution sum up to zero.
}
\end{figure}

We first discuss the Gilbert damping in the collinear case, 
i.e., we set $\magdir(\vn{r})=\hat{\vn{e}}_{z}$ in 
Eq.~\eqref{eq_rashba_model_onedim}.
The $xx$ component of the Gilbert damping is shown 
in  Fig.~\ref{fig_damping_onedim_rashba_collinear_vs_scattering_nosoi} 
as a function of scattering strength $U$
for the
following model parameters: exchange 
splitting $\Delta V=$1eV,
Fermi energy $\mathcal{E}_{\rm F}=2.72$eV
and SOI strength $\alpha^{\rm R}=0$.
All three contributions are individually non-zero,
but the contribution from the 
RR-vertex correction (Eq.~\eqref{eq_s_rr_vert}) 
is much smaller than the one from the 
AR-vertex correction (Eq.~\eqref{eq_s_ra_vert}) 
and much smaller than the intrinsic contribution (Eq.~\eqref{eq_s_int}).
However, in this case the total damping is zero, because a non-zero damping
in periodic crystals with collinear magnetization is only possible 
when SOI is present~\cite{gilbert_damping_model_tests}.

\begin{figure}
\includegraphics[width=\linewidth]{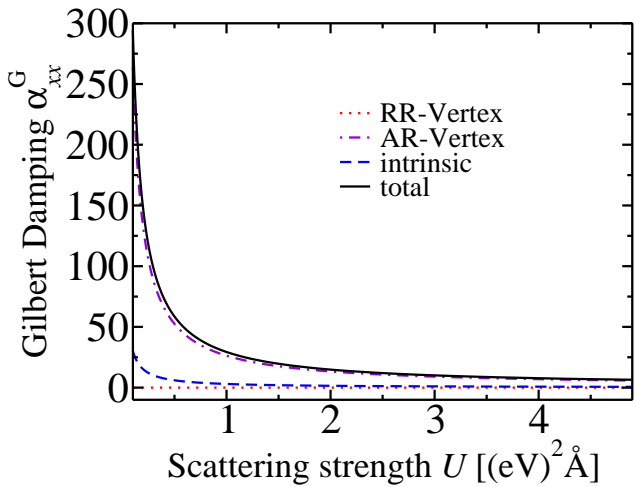}
\caption{\label{fig_damping_onedim_rashba_collinear_vs_scattering}
Gilbert damping $\alpha^{\rm G}_{xx}$ vs.\ 
scattering strength $U$ 
in the one-dimensional Rashba model with SOI.
}
\end{figure}

In  Fig.~\ref{fig_damping_onedim_rashba_collinear_vs_scattering} 
we show the $xx$ component of the 
Gilbert damping $\alpha^{\rm G}_{xx}$ as a function of 
scattering strength $U$ for the
model parameters $\Delta V=1$eV,
$\mathcal{E}_{\rm F}=2.72$eV
and $\alpha^{\rm R}=2$eV\AA.
The dominant contribution is the AR-vertex correction.
The damping as obtained based on Eq.~\eqref{eq_lambda_torque} 
diverges like $1/U$ in the limit $U\rightarrow 0$, i.e.,
proportional to the relaxation time $\tau$~\cite{gilbert_damping_model_tests}.
However, once the relaxation time $\tau$ is larger than the inverse frequency of the
magnetization dynamics the dc-limit $\omega\rightarrow 0$ in Eq.~\eqref{eq_lambda_torque}
is not appropriate and $\omega>0$ needs to be used. 
It has been shown that the Gilbert damping is not infinite in the
ballistic limit $\tau\rightarrow\infty$ 
when $\omega>0$~\cite{absence_intraband_gilbert_damping,breakdown_adiabatic_approach_damping}.
In the one-dimensional Rashba model the effective magnetic
field exerted by SOI on the electron spins points in $y$ direction. Since
a magnetic field along $y$ direction cannot lead to a torque in $y$ direction
the $yy$ component of the Gilbert 
damping $\alpha^{\rm G}_{yy}$ is zero (not shown in the Figure).

\begin{figure}
\includegraphics[width=\linewidth]{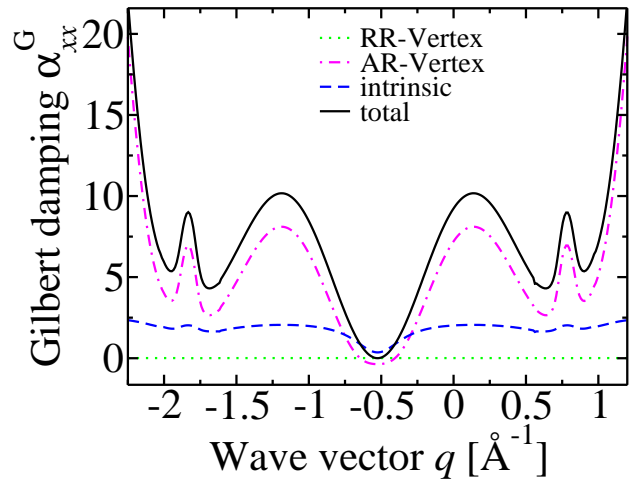}
\caption{\label{fig_rashba_onedim_damping11_vs_qvec}
Gilbert damping $\alpha^{\rm G}_{xx}$ 
vs.\ spin spiral wave number $q$ 
in the one-dimensional
Rashba model.
}
\end{figure}

Next, we discuss the Gilbert damping in the noncollinear case.
In Fig.~\ref{fig_rashba_onedim_damping11_vs_qvec}
we plot the $xx$ component
of the Gilbert damping as a function of spin spiral wave number $q$
for the model 
parameters $\Delta V=1$eV, $\mathcal{E}_{\rm F}=1.36$eV, $\alpha^{\rm R}=2$eV\AA,
and the scattering strength $U=0.98$(eV)$^{2}$\AA. The curves are symmetric 
around $q=-2 m_e \alpha^{\rm R}/\hbar^2$, because the
damping is determined by the effective SOI defined in 
Eq.~\eqref{eq_effective_soi}. 
At $q=-2 m_e \alpha^{\rm R}/\hbar^2$ the effective SOI is zero and therefore
the total damping is zero as well.
The damping at wave number $q$ differs from 
the one at wave number $-q$, i.e., 
\textit{the damping is chiral in the Rashba model}.
Around the 
point $q=-2 m_e \alpha^{\rm R}/\hbar^2$ the damping is described by a 
quadratic parabola at first.
In the regions -2\AA$^{-1}<q<$-1.2\AA$^{-1}$ 
and 0.2\AA$^{-1}<q<$1\AA$^{-1}$
this trend is interrupted by a W-shape behaviour.
In the quadratic parabola region the lowest energy band crosses the
Fermi energy twice. As shown in Fig.~\ref{fig_bands_onedim} the lowest
band has a local maximum at $q=0$. In the W-shape region this local
maximum shifts upwards, approaches the Fermi level and finally passes it
such that the lowest energy band crosses the Fermi level four times.
This transition in the band structure leads to oscillations in the density of
states, which results in the W-shape behaviour of the Gilbert damping.

Since the damping is determined by the effective SOI, we can use
Fig.~\ref{fig_rashba_onedim_damping11_vs_qvec} to draw conclusions
about the damping in the noncollinear case with $\alpha^{\rm R}=0$: We only
need to shift all curves in Fig.~\ref{fig_rashba_onedim_damping11_vs_qvec}
to the right such that they are symmetric around $q=0$ and shift the Fermi energy. 
Thus, for $\alpha^{\rm R}=0$ the Gilbert damping does not vanish if $q \ne 0$.
Since for $\alpha^{\rm R}=0$ angular momentum transfer from the electronic system
to the lattice is not possible, the damping is purely nonlocal in this case, i.e., angular
momentum is interchanged between electrons at different positions.
This means that for a volume in which the magnetization
of the spin-spiral in Eq.~\eqref{eq_spin_spiral_cycloid} 
performs exactly one revolution between
one end of the volume and the other end the total angular momentum
change associated with the damping is zero, because the angular
momentum
is simply redistributed within this volume and there is no net change of the
angular momentum. A substantial contribution of nonlocal damping 
has also been predicted  for domain walls in 
permalloy~\cite{gilbert_damping_noncollinear_ferromagnets_PhysRevLett.113.266603}.

\begin{figure}
\includegraphics[width=\linewidth]{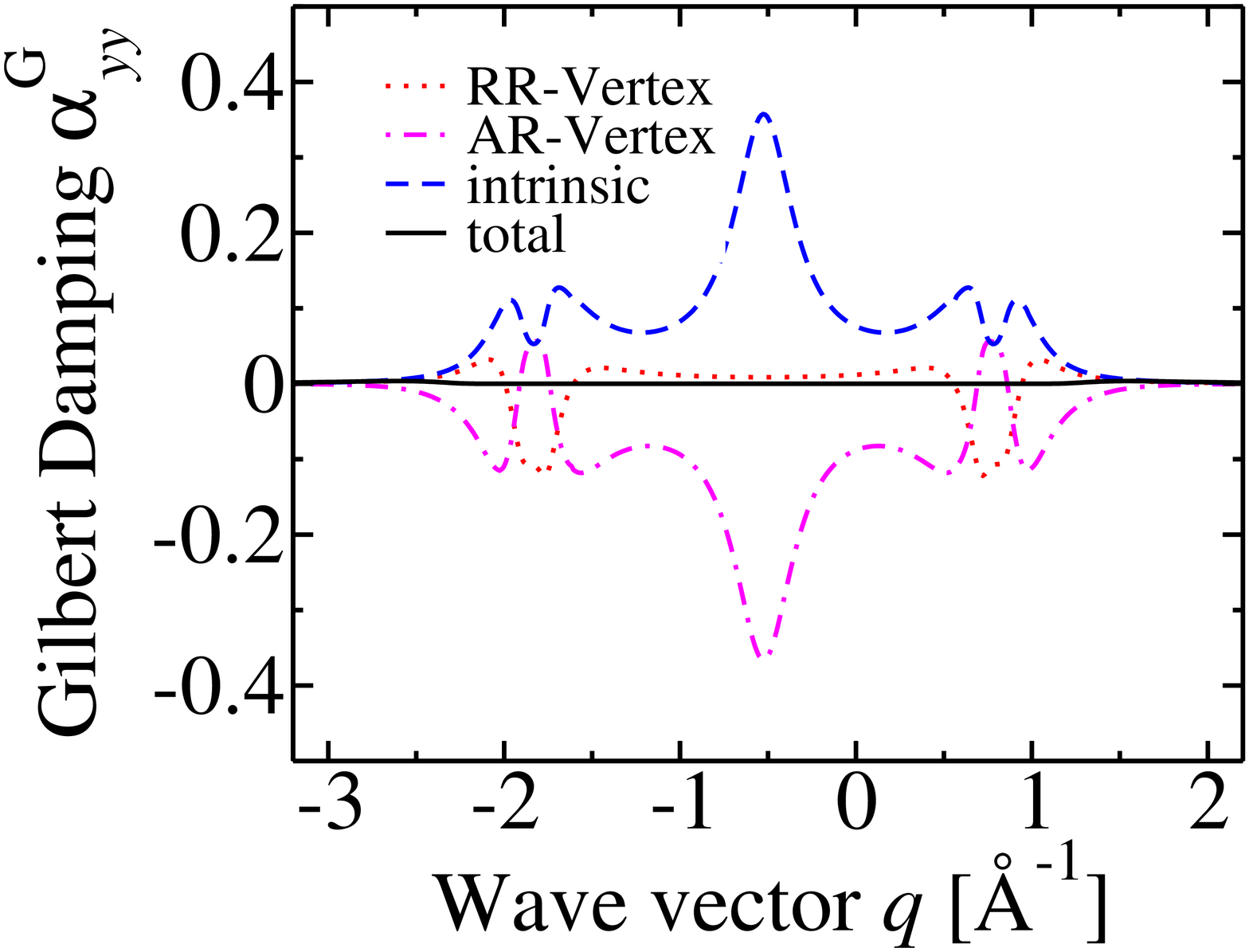}
\caption{\label{fig_rashba_onedim_damping22_vs_qvec}
Gilbert damping $\alpha^{\rm G}_{yy}$ 
vs.\ spin spiral wave number $q$ 
in the one-dimensional
Rashba model.
}
\end{figure}

In Fig.~\ref{fig_rashba_onedim_damping22_vs_qvec}
we plot the $yy$ component
of the Gilbert damping as a function of spin spiral wave number $q$
for the model 
parameters $\Delta V=1$eV, $\mathcal{E}_{\rm F}=1.36$eV, $\alpha^{\rm R}=2$eV\AA,
and the scattering 
strength $U=0.98$(eV)$^2$\AA. 
The total damping is zero in this case. This
can be understood from the symmetry properties of the 
one-dimensional Rashba 
Hamiltonian, Eq.~\eqref{eq_rashba_model_onedim}: Since this
Hamiltonian is invariant when both $\vn{\sigma}$ 
and $\hat{\vn{M}}$ are rotated around the $y$ axis,
the damping coefficient $\alpha^{\rm G}_{yy}$ does not depend on the
position within the cycloidal spin spiral of Eq.~\eqref{eq_spin_spiral_cycloid}.
Therefore, nonlocal damping is not possible in this case and
$\alpha^{\rm G}_{yy}$ has to be zero when $\alpha^{\rm R}=0$.
It remains to be shown that $\alpha^{\rm G}_{yy}=0$ also 
for $\alpha^{\rm R} \ne 0$.
However, this follows directly from the observation that the damping
is determined by the effective SOI, Eq.~\eqref{eq_effective_soi}, meaning
that any case with $q\ne 0$ and $\alpha^{\rm R}\ne 0$ can 
always be mapped onto a case with $q\ne 0$ and $\alpha^{\rm R}=0$.
As an alternative argumentation we can also invoke the finding
discussed above that $\alpha^{\rm G}_{yy}=0$ in the collinear case. Since
the damping is determined by the effective SOI, it follows that
$\alpha^{\rm G}_{yy}=0$ also in the noncollinear case.

\subsection{Current-induced torques}
\label{sec_results_cit}
\begin{figure}
\includegraphics[width=\linewidth]{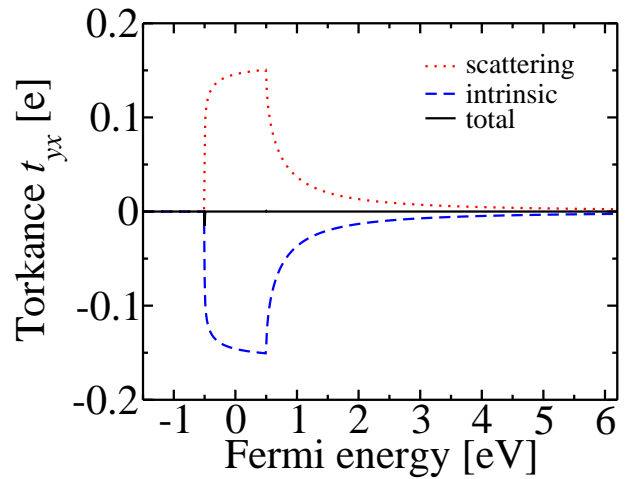}
\caption{\label{fig_onedim_sot_vs_fermi}
Torkance $t_{yx}$ 
vs.\ Fermi energy $\mathcal{E}_{\rm F}$ 
in the one-dimensional
Rashba model.
}
\end{figure}

We first discuss the $yx$ component of the torkance.
In Fig.~\ref{fig_onedim_sot_vs_fermi} we show the torkance $t_{yx}$
as a function of the Fermi energy $\mathcal{E}_{\rm F}$ for the 
model parameters $\Delta V=1$eV 
and $\alpha^{\rm R}=2$eV\AA\, when the
magnetization is collinear and points in $z$ direction.
We specify the torkance in units of the positive elementary charge $e$, which
is a convenient choice for the one-dimensional Rashba model. When the 
torkance is multiplied with the electric field, we obtain the torque per
length (see Eq.~\eqref{eq_torque_and_torkance} 
and Ref.~\cite{note_on_units}).
Since the effective magnetic field from SOI points in $y$ direction,
it cannot give rise to a torque in $y$ direction and consequently
the total $t_{yx}$ is zero.
Interestingly, the intrinsic and
scattering
contributions are individually nonzero.
The intrinsic contribution is nonzero, because the electric field accelerates
the electrons such that $\hbar \dot{k}_{x}=-e E_{x}$. Therefore,
the effective magnetic field $B^{\rm SOI}_{y}=\alpha^{\rm R}k_{x}/\mu_{\rm B}$
changes as well, i.e., $\dot{B}^{\rm SOI}_{y}=\alpha^{\rm R}\dot{k}_{x}/\mu_{\rm B}=-\alpha^{\rm R}E_{x}e/(\hbar \mu_{\rm B})$.
Consequently, the electron spin is no longer aligned with the total effective magnetic
field (the effective magnetic field resulting from both SOI and from the exchange 
splitting $\Delta V$), when an electric field is applied. 
While the total effective magnetic field lies in 
the $yz$ plane, the electron spin acquires an $x$ component, because it
precesses around the total effective magnetic field, with which it is not
aligned due to the applied electric field~\cite{sot_kurebayashi}. 
The $x$ component of the
spin density results in a torque in $y$ direction, which is the
reason why the intrinsic contribution to $t_{yx}$ is nonzero.
The scattering contribution to $t_{yx}$ cancels the intrinsic contribution 
such that the total $t_{yx}$ is zero and angular momentum conservation
is satisfied.  

Using the concept of effective SOI, Eq.~\eqref{eq_effective_soi}, we 
conclude that $t_{yx}$ is also zero for the noncollinear spin-spiral
described by Eq.~\eqref{eq_spin_spiral_cycloid}. Thus, both the
$y$ component of the spin-orbit
torque and the nonadiabatic torque are zero for the one-dimensional
Rashba model.

\begin{figure}
\includegraphics[width=\linewidth]{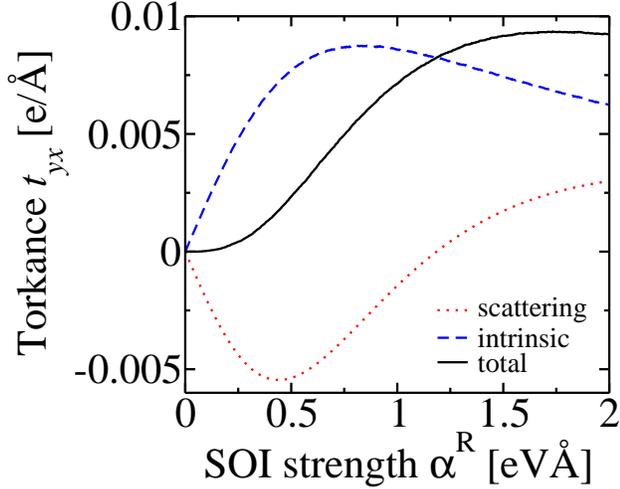}
\caption{\label{fig_rashba_tovelocorr_twodim_vs_alpha_yx}
Nonadiabatic torkance $t_{yx}$ vs.\ SOI 
parameter $\alpha^{\rm R}$ in the two-dimensional
Rashba model.  
}
\end{figure}

To show that $t_{yx}=0$ is a peculiarity of the one-dimensional Rashba model,
we plot in Fig.~\ref{fig_rashba_tovelocorr_twodim_vs_alpha_yx}
the torkance $t_{yx}$ in the two-dimensional Rashba model.
The intrinsic and scattering contributions depend
linearly on $\alpha^{\rm R}$ for small values of $\alpha^{\rm R}$, but the slopes
are opposite such that the total $t_{yx}$ is zero for sufficiently small $\alpha^{\rm R}$.
However, for larger values of $\alpha^{\rm R}$ the intrinsic and scattering contributions
do not cancel each other and therefore  
the total $t_{yx}$ becomes nonzero, in contrast to 
the one-dimensional Rashba model, where $t_{yx}=0$ even for large $\alpha^{\rm R}$.
Several previous works determined the part of $t_{yx}$ that is
proportional to $\alpha^{\rm R}$ 
in the two-dimensional Rashba model and 
found it to be zero~\cite{sots_2d_rashba_ferromagnets_titov,microscopic_theory_sot}
for scalar disorder,
which is consistent with our finding that the linear slopes of the
intrinsic and scattering contributions to $t_{yx}$ are opposite for small $\alpha^{\rm R}$.   

\begin{figure}
\includegraphics[width=\linewidth]{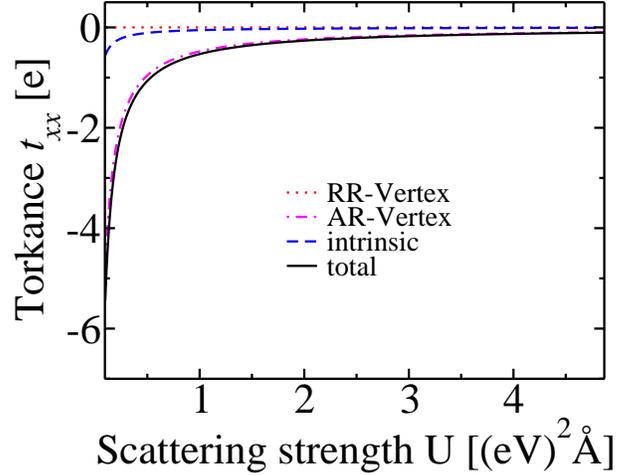}
\caption{\label{fig_onedim_stt_vs_sigma}
Torkance $t_{xx}$ 
vs.\ scattering strength $U$ 
in the one-dimensional
Rashba model.
}
\end{figure}

Next, we discuss the $xx$ component of the torkance in the collinear
case ($\hat{\vn{M}}=\hat{\vn{e}}_{z}$).
In Fig.~\ref{fig_onedim_stt_vs_sigma} we plot the
torkance $t_{xx}$ 
vs.\ scattering strength $U$ 
in the one-dimensional
Rashba model for the
parameters $\Delta V=1$eV, $\mathcal{E}_{\rm F}=2.72$eV
and $\alpha^{\rm R}=2$eV\AA. 
The dominant contribution is the AR-type vertex 
correction (see Eq.~\eqref{eq_tovelo_odd_vert_ra}).
$t_{xx}$ diverges like $1/U$ in the limit $U\rightarrow 0$
as expected for the odd torque in metallic systems~\cite{ibcsoit}.

\begin{figure}
\includegraphics[width=\linewidth]{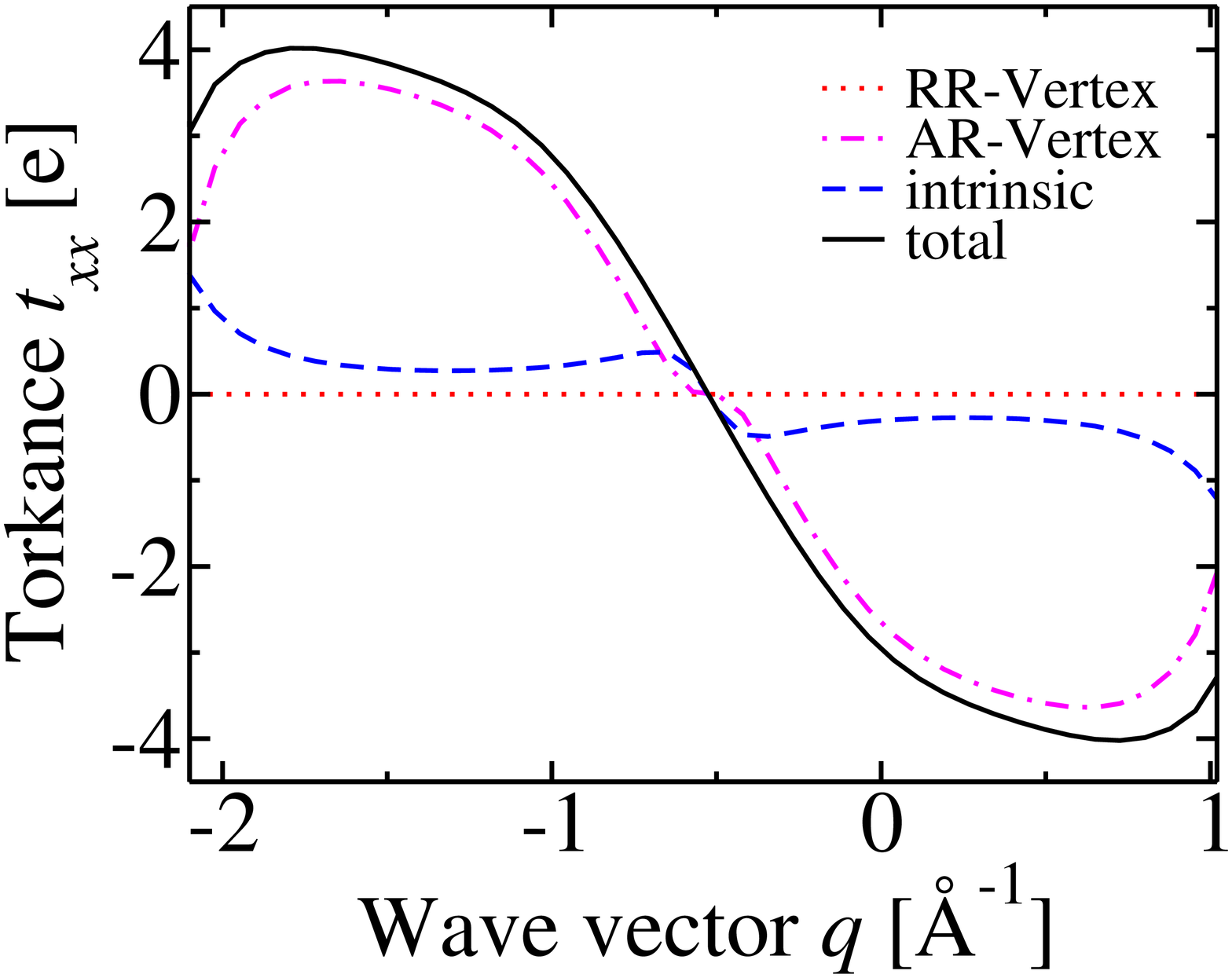}
\caption{\label{fig_onedim_stt_vs_qvec}
Torkance $t_{xx}$ 
vs.\ wave vector $q$ 
in the one-dimensional
Rashba model with SOI.
}
\end{figure}

\begin{figure}
\includegraphics[width=\linewidth]{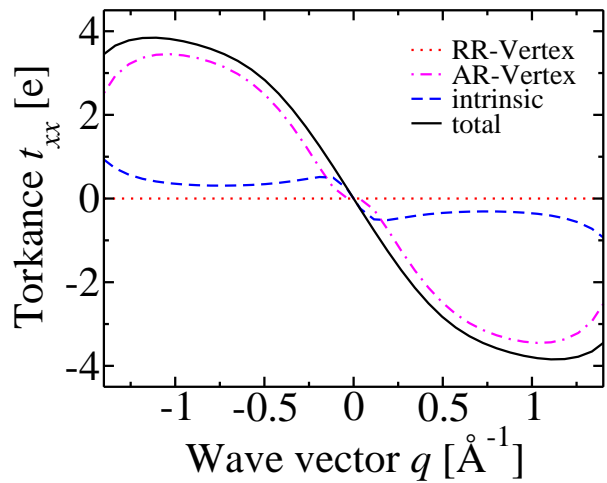}
\caption{\label{fig_onedim_stt_vs_qvec_nosoi}
Torkance $t_{xx}$ 
vs.\ wave vector $q$ 
in the one-dimensional
Rashba model without SOI.
}
\end{figure}

In Fig.~\ref{fig_onedim_stt_vs_qvec} 
and Fig.~\ref{fig_onedim_stt_vs_qvec_nosoi} 
we plot $t_{xx}$
as a function of spin-spiral wave number $q$ for the
model parameters $\Delta V=1$eV, $\mathcal{E}_{\rm F}=2.72$eV
and $U=0.18$(eV)$^2$\AA.
In  Fig.~\ref{fig_onedim_stt_vs_qvec} the case 
with $\alpha^{\rm R}=2$eV\AA\, is shown, while 
Fig.~\ref{fig_onedim_stt_vs_qvec_nosoi} illustrates the
case with $\alpha^{\rm R}=0$.
In the case $\alpha^{\rm R}=0$ the torkance $t_{xx}$ describes the
spin-transfer torque (STT).
In the case $\alpha^{\rm R}\ne 0$ the torkance $t_{xx}$ is the sum of
contributions from STT
and spin-orbit torque (SOT).
The curves with $\alpha^{\rm R}=0$ and $\alpha^{\rm R}\ne 0$ are 
essentially related by a shift
of $\Delta q=-2 m_{e} \alpha^{\rm R}/\hbar^2$, which 
can be understood based on the
concept of the effective SOI, Eq.~\eqref{eq_effective_soi}.
Thus, in the special case of the one-dimensional
Rashba model STT and SOT are strongly related.

\section{Summary}
\label{sec_summary}
We study chiral damping, chiral gyromagnetism and
current-induced torques in the one-dimensional 
Rashba model with an additional N\'eel-type
noncollinear magnetic exchange field.
In order to describe scattering effects we use a Gaussian scalar
disorder model. Scattering contributions are generally
important in the one-dimensional Rashba model with the
exception of the gyromagnetic ratio in the collinear case with zero SOI,
where the scattering corrections vanish in the clean limit.
In the one-dimensional Rashba model SOI and 
noncollinearity can be combined into an effective SOI.
Using the concept of effective SOI, results for the magnetically collinear one-dimensional
Rashba model can be used to predict the behaviour in the noncollinear 
case. In the noncollinear Rashba model the Gilbert damping is nonlocal
and does not vanish for zero SOI. The scattering corrections
tend to stabilize the gyromagnetic ratio in the
one-dimensional Rashba model at its nonrelativistic value.
Both the Gilbert damping and the gyromagnetic ratio are chiral for
nonzero SOI strength.
The antidamping-like spin-orbit torque and the nonadiabatic 
torque for N\'eel-type spin-spirals are zero
in the one-dimensional Rashba model, while the
antidamping-like spin-orbit torque is nonzero in the two-dimensional 
Rashba model for sufficiently large SOI-strength.

\bibliography{chigyromag}

\begin{thebibliography}{54}
\bibitem[{not()}]{note_on_units}
\bibinfo{note}{In Ref.~\cite{ibcsoit} and Ref.~\cite{invsot} we consider the
  torque per unit cell and use therefore charge times length as unit of
  torkance. In this work we consider instead the torque per volume.
  Consequently, the unit of torkance used in this work is charge times length
  per volume. In the one-dimensional Rashba model the volume is given by the
  length and consequently the unit of torkance is charge.}
\end{thebibliography}

\end{document}